\documentclass[manuscript,screen]{acmart}
\usepackage{parcolumns}
\usepackage[colorinlistoftodos,textwidth=15mm,textsize=tiny,obeyFinal]{todonotes}
\usepackage{subcaption}
\usepackage{mathtools} 
\usepackage{color}
\usepackage{textcomp}
\usepackage{etoolbox}
\usepackage{graphicx}
\usepackage{colortbl}
\usepackage{framed}
\usepackage{tabularx}
\usepackage[linesnumbered,ruled, vlined]{algorithm2e}
\usepackage{multirow,bigdelim}
\usepackage{enumitem}
\usepackage{balance}
\usepackage{url}
\usepackage[utf8]{inputenc}
\usepackage{booktabs}
\usepackage{threeparttable}
\usepackage{multirow}
\usepackage{adjustbox}
\usepackage{xcolor}
\usepackage{listings}
\usepackage{tcolorbox}
\tcbuselibrary{skins, breakable}
\usepackage{url}
\usepackage{tikz}
\usetikzlibrary{decorations,calligraphy}
\usepackage{fancyvrb}
\tcbuselibrary{breakable} % 支持跨页显示

\usepackage[english]{babel}
\usepackage{amsthm}

\usepackage[frozencache,cachedir=minted-cache]{minted}
\setminted{
  linenos,
  breaklines,
  autogobble,
  fontsize=\small,
  tabsize=4
}
\AtBeginDocument{%
  }

\setcopyright{acmlicensed}
\copyrightyear{2026}
\acmYear{2026}
\acmDOI{XXXXXXX.XXXXXXX}
\acmISBN{}

\begin{document}

%%
%% The "title" command has an optional parameter,
%% allowing the author to define a "short title" to be used in page headers.
% \title{Universal Testing of Static Application Security Testing Tools via LLM with Dynamic Feedback}
% \title{Multi-language Metamorphic Testing of Static Application Security Testing Tools via Semantically Diverse Program Transformations}
\title{Combining LLM and Dynamic Analysis for Universal Testing of Static Application Security Testing Tools}
% \title{Universal Testing of Static Application Security Testing Tools via LLM-Assisted Dynamic Analysis}

%%
%% The "author" command and its associated commands are used to define
%% the authors and their affiliations.
%% Of note is the shared affiliation of the first two authors, and the
%% "authornote" and "authornotemark" commands
%% used to denote shared contribution to the research.
% \author{Ben Trovato}
% \authornote{Both authors contributed equally to this research.}
% \email{trovato@corporation.com}
% \orcid{1234-5678-9012}
% \author{G.K.M. Tobin}
% \authornotemark[1]
% \email{webmaster@marysville-ohio.com}
% \affiliation{%
%   \institution{Institute for Clarity in Documentation}
%   \city{Dublin}
%   \state{Ohio}
%   \country{USA}
% }

\author{Huaien Zhang}
\affiliation{
  \institution{The University of Hong Kong and the Hong Kong Polytechnic University}
  \city{Hong Kong}
  \country{China}}
\email{huaien@hku.hk}

\author{Yu Pei}
\affiliation{
  \institution{The Hong Kong Polytechnic University}
  \city{Hong Kong}
  \country{China}
}
\email{csypei@comp.polyu.edu.hk}

\author{Hao Chen}
\affiliation{
 \institution{The University of Hong Kong}
 \city{Hong Kong}
 \country{China}
}
\email{chenho@hku.hk}

\author{Shin Hwei Tan}
\affiliation{
 \institution{Concordia University}
 \city{Montreal}
 \country{Canada}}
\email{shinhwei.tan@concordia.ca}

%%
%% By default, the full list of authors will be used in the page
%% headers. Often, this list is too long, and will overlap
%% other information printed in the page headers. This command allows
%% the author to define a more concise list
%% of authors' names for this purpose.
\renewcommand{\shortauthors}{Huaien et al.}

\definecolor{main-color}{rgb}{0.6627, 0.7176, 0.7764}
\definecolor{string-color}{rgb}{0.3333, 0.5254, 0.345}
\definecolor{key-color}{rgb}{0.8, 0.47, 0.196}

% \newtcolorbox{systemprompt}{
%     colback=red!5!white,
%     colframe=red!70!black,
%     title=System Prompt,
%     fonttitle=\bfseries,
%     breakable,
%     enhanced,
% }

\newtcolorbox{systemprompt}{
    width=\textwidth,        % 明确设置宽度
    colback=red!5!white,
    colframe=red!70!black,
    title=System Prompt,
    % fonttitle=\bfseries,
    enhanced,
    % sharp corners,          % 直角边框
    % 对齐设置
    halign=left,            % 内容区左对齐
    % halign title=left,    % 标题居中对齐（可选）
    valign=top,             % 内容区顶部对齐
    before upper={\parindent0pt\raggedright}, % 强制左对齐且无缩进
    % before upper={\parindent0pt}, % 确保无缩进
}

% \newtcolorbox{userprompt}{
%     width=\linewidth, % 关键：设置框体宽度为行宽
%     colback=green!5!white,
%     colframe=green!70!black,
%     title=User Prompt,
%     fonttitle=\bfseries,
%     breakable,
%     enhanced,
% }
\newtcolorbox{userpromptwithtitle}{
    width=\textwidth,        % 明确设置宽度
    colback=green!5!white,
    colframe=green!70!black,
    title=User Prompt,
    fonttitle=\bfseries,
    enhanced,
    % sharp corners,          % 直角边框
    % 对齐设置
    halign=left,            % 内容区左对齐
    halign title=left,    % 标题居中对齐（可选）
    valign=top,             % 内容区顶部对齐
    before upper={\parindent0pt\raggedright}, % 强制左对齐且无缩进
    % before upper={\parindent0pt}, % 确保无缩进
}

\newtcolorbox{userprompt}{
    width=\textwidth,        % 明确设置宽度
    colback=green!5!white,
    colframe=green!70!black,
    % fonttitle=\bfseries,
    enhanced,
    % sharp corners,          % 直角边框
    % 对齐设置
    halign=left,            % 内容区左对齐
    % halign title=left,    % 标题居中对齐（可选）
    valign=top,             % 内容区顶部对齐
    before upper={\parindent0pt\raggedright}, % 强制左对齐且无缩进
    % before upper={\parindent0pt}, % 确保无缩进
}

% \newtcolorbox{userprompt}{
%     colback=green!5!white,
%     colframe=green!70!black,
%     title=User Prompt,
%     fonttitle=\bfseries,
%     breakable,
%     enhanced,
% }

\newtcolorbox{userprompt2}{
    enhanced, % 关键：启用增强引擎
    width=\textwidth, % 确保宽度一致
    arc=0pt, % 直角边框
    outer arc=0pt,
    boxrule=0.5pt, % 边框粗细
    colback=green!5!white, % 内容区背景色
    colframe=green!70!black, % 边框颜色
    colbacktitle=green!70!black, % 标题栏背景色（关键参数）
    coltitle=white, % 标题文字颜色（关键参数）
    fonttitle=\bfseries\sffamily, % 标题字体
    title=User Prompt, % 标题文本
    titlerule=0pt, % 移除默认标题下线
    attach title to upper=% 创建自定义分隔线
        {\linebreak[0]\vspace{2pt}\rule[0pt]{\linewidth}{0.4pt}\vspace{2pt}},
    breakable,
    left=6pt, % 内容区内边距
    right=6pt,
    top=4pt,
    bottom=4pt,
}

\newtcolorbox{aiprompt}{
    colback=purple!5!white,
    colframe=purple!70!black,
    title=AI Response,
    fonttitle=\bfseries,
    breakable,
    enhanced,
}

\newtcolorbox{promptbox}{
    colback=gray!5,      % 背景色（浅灰）
    colframe=gray!50,     % 边框色
    arc=3pt,              % 圆角弧度
    boxsep=5pt,           % 内边距
    left=6pt, right=6pt, % 左右边距
    breakable,            % 允许跨页
    fontupper=\ttfamily,  % 等宽字体（关键！）
    title={\small\textbf{Prompt}}, % 标题（可选）
}

\definecolor{GrayCodeBlock}{RGB}{241,241,241}
\definecolor{BlackText}{RGB}{110,107,94}
\definecolor{RedTypename}{RGB}{182,86,17}
\definecolor{GreenString}{RGB}{96,172,57}
\definecolor{PurpleKeyword}{RGB}{184,84,212}
\definecolor{GrayComment}{RGB}{170,170,170}
\definecolor{GoldDocumentation}{RGB}{180,165,45}
\lstdefinelanguage{rust}
{
    columns=fullflexible,
    keepspaces=true,
    frame=single,
    framesep=0pt,
    framerule=0pt,
    framexleftmargin=4pt,
    framexrightmargin=4pt,
    framextopmargin=5pt,
    framexbottommargin=3pt,
    xleftmargin=4pt,
    xrightmargin=4pt,
    backgroundcolor=\color{GrayCodeBlock},
    basicstyle=\ttfamily\color{BlackText},
    keywords={
        true,false,
        unsafe,async,await,move,
        use,pub,crate,super,self,mod,
        struct,enum,fn,const,static,let,mut,ref,type,impl,dyn,trait,where,as,
        break,continue,if,else,while,for,loop,match,return,yield,in
    },
    keywordstyle=\color{PurpleKeyword},
    ndkeywords={
        bool,u8,u16,u32,u64,u128,i8,i16,i32,i64,i128,char,str,
        Self,Option,Some,None,Result,Ok,Err,String,Box,Vec,Rc,Arc,Cell,RefCell,HashMap,BTreeMap,
        macro_rules
    },
    ndkeywordstyle=\color{RedTypename},
    comment=[l][\color{GrayComment}\slshape]{//},
    morecomment=[s][\color{GrayComment}\slshape]{/*}{*/},
    morecomment=[l][\color{GoldDocumentation}\slshape]{///},
    morecomment=[s][\color{GoldDocumentation}\slshape]{/*!}{*/},
    morecomment=[l][\color{GoldDocumentation}\slshape]{//!},
    morecomment=[s][\color{RedTypename}]{\#![}{]},
    morecomment=[s][\color{RedTypename}]{\#[}{]},
    stringstyle=\color{GreenString},
    string=[b]"
}

\lstdefinestyle{mystyle}
{
    language = SQL,
    basicstyle = {\ttfamily \color{main-color}},
    stringstyle = {\color{string-color}},
    keywordstyle = {\color{key-color}},
    keywordstyle = [2]{\color{lime}},
    keywordstyle = [3]{\color{yellow}},
    keywordstyle = [4]{\color{teal}},
    morekeywords = [3]{<<, >>},
    morekeywords = [4]{++},
    basicstyle=\ttfamily\footnotesize,
    commentstyle=\color{blue}\ttfamily,
    morecomment=[f][\lstbg{red!20}]-,
    morecomment=[f][\lstbg{green!20}]+,
    morecomment=[f][\textit]{@@},
    numbers=left,
    texcl=false
}

\definecolor{pgrey}{rgb}{0.46,0.45,0.48}

\lstdefinestyle{javastyle}
{
    language = Java,
    basicstyle = {\ttfamily \color{main-color}},
    stringstyle = {\color{string-color}},
    keywordstyle = {\color{key-color}},
    keywordstyle = [2]{\color{lime}},
    keywordstyle = [3]{\color{yellow}},
    keywordstyle = [4]{\color{teal}},
    morekeywords = [3]{<<, >>},
    morekeywords = [4]{++},
    morekeywords = {List, Object},
    basicstyle=\ttfamily\scriptsize,
    commentstyle=\color{blue}\ttfamily,
    morecomment=[f][\lstbg{red!20}]-,
    morecomment=[f][\lstbg{green!20}]+,
    morecomment=[f][\textit]{@@},
    numbers=left,
    xleftmargin=4mm,
    numbersep=7pt,
    numberstyle=\scriptsize,
    frame=none,
    texcl=false,
    moredelim=[il][\textcolor{pgrey}]{\$\$},
    moredelim=[is][\textcolor{pgrey}]{\%\%}{\%\%}
}

\definecolor{grey}{rgb}{0.7,0.7,0.7}

%diff
\newcommand{\lstbg}[3][0pt]{{\fboxsep#1\colorbox{#2}{\strut #3}}}
\lstdefinelanguage{diff}{
  basicstyle=\ttfamily\footnotesize,,
  morecomment=[f][\lstbg{red!20}]-,
  morecomment=[f][\lstbg{green!20}]+,
  morecomment=[f][\textit]{@@},
  texcl=false
}

\newtcolorbox{definitionA}[1][]{
  enhanced,
  colback=white,
  colframe=blue!60!black,
  arc=3pt,
  boxrule=0.8pt,
  left=6pt,
  right=6pt,
  top=8pt,
  bottom=8pt,
  fonttitle=\bfseries,
  title=Definition,
  #1
}

\newtcolorbox{definitionC}[1][]{
  enhanced,
  colback=gray!3!white,
  colframe=gray!60!black,
  arc=2pt,
  boxrule=0.5pt,
  left=8pt,
  right=8pt,
  top=10pt,
  bottom=10pt,
  fonttitle=\bfseries,
  coltitle=black,
  title=Definition,
  #1
}

\lstdefinelanguage{Rust}{
  keywords={fn, let, mut, unsafe, vec, pub, use, mod, struct, enum, impl, trait, match, if, else, while, for, in, return, as},
  sensitive=true,
  comment=[l]{//},
  morecomment=[s]{/*}{*/},
  string=[b]"
}

\lstset{
    language=Rust,
    basicstyle=\ttfamily\small,
    breaklines=true,
    numbers=left,           % 显示行号（对应你原先的行号需求）
    numberstyle=\tiny\color{gray},
    numbersep=4pt,          % 行号与代码间距
    xleftmargin=1em,        % 左侧缩进
    commentstyle=\color{gray}\itshape,
    keywordstyle=\color{blue}\bfseries,
    stringstyle=\color{orange},
    showstringspaces=false,
    frame=none              % 如果需要边框可以改成 frame=single
}

\def\sys{{\sc LLM4SAST}\xspace}
\def\sysb{{\sc \textbf{LLM4SAST}}\xspace}

\newcommand{\yp}[1]{\todo[color=green!40]{#1}}
\newcommand{\zhe}[1]{\todo[color=blue!30]{zhe:#1}}

\newcommand{\modify}[1]{\textcolor{blue}{#1}}
\newcommand{\allissuecnt}{308}
\newcommand{\issuecnt}{246}
\newcommand{\faultcnt}{66}
\newcommand{\fixedfaultcnt}{32}
\newcommand{\fpcnt}{seven}
\newcommand{\rootcausecnt}{six}
\newcommand{\symptomcnt}{four}
\newcommand{\fixstrategycnt}{seven}
\newcommand{\findingcnt}{ten}
\newcommand{\clustercnt}{102}

\newcommand{\numtool}{six}
\newcommand{\lockbud}{LockBud\xspace}
\newcommand{\semgrep}{Semgrep\xspace}
\newcommand{\rudra}{Rudra\xspace}
\newcommand{\prusti}{Prusti\xspace}
\newcommand{\miri}{MIRI\xspace}
\newcommand{\sonar}{SonarQube\xspace}
\newcommand{\pmd}{PMD\xspace}
\newcommand{\sbugs}{SpotBugs\xspace}
\newcommand{\cstyle}{CheckStyle\xspace}
\newcommand{\infer}{Infer\xspace}
\newcommand{\soot}{Soot\xspace}
\newcommand{\eprone}{Error-prone\xspace}

\begin{abstract}
Static application security testing (SAST) tools help detect potential software vulnerabilities, but they themselves contain faults undermining reliability and usability.
Automated testing can address these issues, yet current techniques suffer from limited program transformation diversity and lack mechanisms to verify generated variants,
leading to reduced fault detection and high manual effort.
We present \sys{}, the first universal framework detecting faults in SAST tools using large language models and dynamic feedback.
Our key insight is leveraging LLMs to understand and transform code in diverse programming languages.
We design novel metamorphic relations with transformations to guide variant generation, using an LLM to produce multi-language variants that satisfy these relations.
To reduce erroneous variants and identify unique faults, we propose two heuristics: dynamic-feedback-driven selection and program-differencing-based clustering.
For each metamorphic relation, a report-centric test oracle detects variants that fail tests and identifies underlying faults by differentially analyzing SAST reports of the seed program and its variants.
\sys{} identified 66 unique faults across six SAST tools, with 32 confirmed or fixed after being reported to developers, demonstrating effectiveness.
The proposed heuristics eliminate 47.8\% of false positives, and LLM4SAST detects 3.4 times more faults while requiring only 39.8\% of the variants compared to existing approaches.
\end{abstract}

%%
%% The code below is generated by the tool at http://dl.acm.org/ccs.cfm.
%% Please copy and paste the code instead of the example below.
%%


%%
%% Keywords. The author(s) should pick words that accurately describe
%% the work being presented. Separate the keywords with commas.

% \received{20 February 2007}
% \received[revised]{12 March 2009}
% \received[accepted]{5 June 2009}

%%
%% This command processes the author and affiliation and title
%% information and builds the first part of the formatted document.
\maketitle

\section{Introduction}
In the contemporary software development lifecycle, the significance of static application security testing (SAST) tools is paramount.
These tools provide a proactive and systematic approach in identifying security vulnerabilities, design flaws, and compliance violations directly within the source code, bytecode, or binary code of an application without dynamic execution.
The core value of SAST lies in its ability to facilitate a ``Shift-Left'' security paradigm, integrating security assessments early in the development phase.
This early detection drastically reduces the cost and effort associated with remediating issues later in the software development lifecycle (SDLC), as fixing a vulnerability during coding is exponentially less expensive than post-deployment.
By automating code analysis against a comprehensive knowledge base of known vulnerability patterns (e.g., SQL injection, cross-site scripting, buffer overflows), 
SAST tools empower developers to write more secure code, foster a culture of security awareness, and ultimately contribute to the delivery of robust and resilient software products in an increasingly threat-prone digital landscape.
Hence, ensuring the correctness and reliability of SAST tools is a critical research topic, as their implementation faults can result in both missed detections of security vulnerabilities and a high false positive rate that severely compromises their usability.

Prior research on quality assurance for SAST tools primarily falls into four paradigms:
(1) differential testing~\cite{2022icpcfinder, sascope},
(2) metamorphic testing~\cite{tdsc2024query, annatester, statfier},
(3) random testing~\cite{cuoq2012testing},
and (4) program generation~\cite{ase2024minotaur}.
The differential testing approach~\cite{2022icpcfinder, sascope} identify potential faults by detecting inconsistencies among common internal analysis techniques, which rely on the common components between two SAST tools.
In contrast, the metamorphic testing approaches~\cite{tdsc2024query, statfier} flag variants that violate predefined metamorphic relations as potentially faulty, while they only focus on a narrow range of program transformations.
Random testing approach~\cite{cuoq2012testing} generates and selects program variants in a purely random manner, leading to inefficient exploration of the search space.
Program generation approach~\cite{ase2024minotaur} relies on SMT solvers to verify safety properties and construct valid benchmarks, but are often highly computationally intensive.
Despite their proven effectiveness, these conventional approaches suffer from several common key limitations.

\noindent\textit{\textbf{C1: Inability to Support Multi-Language SAST Tools.}}
Existing research efforts are predominantly designed for single programming languages.
For instance, Statifier~\cite{statfier} and SAScope~\cite{sascope} target Java, while FuzzSA~\cite{fse2024sabugs} focuses on C++.
Although the underlying methodologies of these approaches exhibit a degree of generalizability, their direct application to other languages remains challenging.
Furthermore, inherent linguistic differences can cause the test oracles and program transformation strategies, which are effective for one language, to fail when applied to another.

\noindent\textit{\textbf{C2: Limitation of Variant Diversity by a Predefined Transformation Set.}}
% \yp{the term ``variant'' doesn't apply to differential testing or program generation. A more general term should be used if we want \sys{} to be viewed as an enhancement to all the three categories. Or, we can position \sys{} as a metamorphic testing approach.}
% \zhe{a bit confused, maybe the old version?}
Traditional research on testing SAST tools via variant generation relies primarily on predefined semantics-preserving transformations~\cite{tdsc2024query, statfier, sascope}.
This results in limited variant diversity, thereby constraining fault detection effectiveness.
As a consequence, such testing often fails to emulate complex real-world scenarios where code behavior is unintentionally modified, leaving gaps in assessing the robustness and correctness of SAST tools.
Moreover, as programming languages evolve with new features and paradigms, SAST tools frequently struggle to keep pace.
For example, Java Records, introduced in 2021, has revealed shortcomings in tools like SonarQube, causing issues such as those reported in~\cite{sonar144097, sonar124154}.
These problems stem from incomplete or inaccurate handling of new syntax and semantics.
Similarly, support for advanced language concepts (such as the ownership system and unsafe blocks in Rust) often remains limited, further restricting analysis capabilities.
% Traditional research for testing SAST tools on variant generation remains predominantly constrained by a predefined set of semantics-preserving program transformations~\cite{tdsc2024query, statfier, sascope}.
% This approach yields variants with limited diversity, ultimately restricting fault detection capabilities.
% Consequently, the testing process may fail to simulate complex real-world scenarios where code behavior is inadvertently altered, leaving gaps in the assessment of the robustness and practical reliability of SAST tools.
% Besides, programming languages continually evolve, with new features and paradigms being regularly introduced.
% SAST tool developers often struggle to keep pace with these updates promptly.
% Therefore, when analyzing code that utilizes modern language features, SAST tools may produce incorrect results or even encounter runtime failures, e.g., Java Records, a feature officially released in 2021, has exposed limitations in tools like SonarQube, leading to issues such as those documented in~\cite{sonar144097, sonar124154}.
% The root cause lies in the incomplete or incorrect modeling of the new syntax and semantics.
% Furthermore, SAST tools often lack robust support for advanced concepts in modern languages like Rust, such as its ownership system, borrow checker, and unsafe code blocks, limiting their analysis capabilities.

\noindent \textit{\textbf{C3: Lack of an Effective Approach for Filtering Invalid Variants.}}
% \yp{Fixed: Is there any existing work that uses LLMs already? If not, this is a challenge for \sys{} rather than the existing approaches. Particularly, when an approach supports only a limited number of transformations, it's not that difficult to validate the transformed programs.}
Given the inherent limitations of testing approaches, for example, an LLM-based approach may suffer from hallucinations~\cite{xu2024hallucination} that produce invalid variants failing to satisfy the constraints specified in the prompt.
While identifying faults with clear oracles (e.g., runtime crashes) is straightforward, prior studies~\cite{tdsc2024query, annatester} indicate that faults in SAST tools encompass not only crashes but also functional issues such as false positives or negatives resulting from flawed code logic.
Existing research work~\cite{xia2024fuzz4all, mut4all} has primarily focused on detecting runtime crashes, which can be directly observed during execution.
In contrast, our work also aims to uncover logic faults, necessitating a validation mechanism to assess whether the generated variants adhere to the constraints.

% \noindent\textit{\textbf{C1: Lack of Support for Evolving Language Features.}}
% Programming languages continually evolve, with new features and paradigms being regularly introduced.
% SAST tool developers often struggle to keep pace with these updates promptly.
% Consequently, when analyzing code that utilizes modern language features, SAST tools may produce incorrect results or even encounter runtime failures. For example, Java Records, a feature officially released in 2021, has exposed limitations in tools like SonarQube, leading to issues such as those documented in ~\cite{sonar144097, sonar124154}.
% The root cause lies in the incomplete or incorrect modeling of the new syntax and semantics.
% Furthermore, SAST tools often lack robust support for advanced concepts in modern languages like Rust, such as its ownership system, borrowing checker\todo{what is borrowing checker? I google and found only borrow checker. please check to ensure the correct terms are used in the paper}, and unsafe blocks, limiting their analysis capabilities.

\noindent \textbf{Our Approach.}
To tackle these challenges, we propose \sys{}, a novel metamorphic testing framework.
It is designed to be universal, capable of testing SAST tools that support multiple programming languages and their evolving features.
Our approach represents a fundamental departure from conventional SAST testing techniques like Statfier~\cite{statfier} and SAST-MT~\cite{tdsc2024query}.
These existing methods employ a limited set of pattern-based program transformations that, resulting in a lack of program diversity, and only target on a single programming language.
The core innovation of our work lies in harnessing large language models as a program transformation engine.
LLMs are uniquely suited for this role due to their extensive pre-training on diverse code corpora spanning multiple programming languages and formal systems.
This training endows them with an intrinsic understanding of complex language semantics and specialized features.
\sys{} capitalizes on this capability by employing an LLM as a universal generation and transformation engine, enabling it to produce semantically-rich, language-aware test inputs that far surpass the diversity achievable by traditional methods.

The input to \sys{} consists of three key components: 
% \yp{fixed: Shouldn't one of the inputs be the SAST to test? Then, input 3 can be produced by applying the SAST on input 1.}
(1) the SAST tool \textit{S} to be tested by \sys{},
% (2) source programs containing at least one issue that \textit{S} can detect, and
(2) an input programs for \textit{S}, and
% \yp{fixed: The introduction mentioned multiple applications of SASTs. Restricting the input to programs with one security vulnerability implies \sys{} has limited applicability. Is this necessary?}
(3) the documentation of \textit{S} that contains vulnerability descriptions.
% (4) the analysis report of \textit{S} for the source programs mentioned in (2) with precise vulnerability locations.
Using these inputs, \sys{} first invokes \textit{S} to analyze the input programs and generate the corresponding reports.
It then constructs input prompts based on predefined templates.
These carefully engineered prompts enable the generation of semantics-diverse program variants specifically designed to expose potential faults in SAST tools.
% To address the issue caused by the drawbacks of LLM (e.g., hallucination~\cite{csur2023hallucination}), which can lead to program variants that violate the constraints stated in the prompts, we developed a dynamic-feedback-driven variant selection technique.
To mitigate issues caused by inherent drawbacks of LLMs, particularly the problem of hallucination~\cite{csur2023hallucination}, which leads to non-compliant program variants, we developed a dynamic-feedback-driven variant selection technique.
% \todo{fixed: we may consider changes the title to include feedback-driven. why Fig 3 (the overview diagram) does not have anything about mutant selection? is variant=mutation? better to use the same phrase}.
By combining runtime status and dynamic analysis results of program variants during test execution, this technique efficiently eliminates invalid variants.
% \todo{fixed: what kind of runtime status? pass/fail? or compilation check? runtime verification? runtime usually refers to during test execution. also runtime status of what (generated variant? input program? differences in the output between input program and generated variant?)?}
Subsequently, the valid LLM-generated variants are submitted to the target SAST tools.
Their analysis reports are then validated using a set of three test oracles to identify potential faults of SAST tools.

\sys{} is designed to overcome the limitations of traditional testing approaches outlined earlier.
In contrast to existing methods, which are often tied to a specific language version or tool (\textbf{C1}), \sys{} can readily adapt to new language features.
This adaptability can be achieved with minimal effort; for instance, simply providing the LLM with documentation or code examples of the new feature is sufficient to guide the generation of relevant test cases.
% \yp{fixed: C2 is about variant diversity. Check again.}
Unlike traditional approaches that support only a narrow range of program transformations, leading to low diversity in program variants (\textbf{C2}), we propose a method utilizing three types of metamorphic relations to guide program transformations from multiple views.
This approach is not limited to semantics-preserving transformations, thereby significantly increasing the semantic diversity of the program variants.
% Unlike traditional approaches that require building a specialized testing framework for each programming language and SAST tool combination (\textbf{C2}), 
% \sys{} employs an LLM as a universal variant generation engine, making it applicable to a wide range of languages.
% \yp{fixed: C3 is about validation oracles. Check again.}
Furthermore, to overcome the inherent limitation of LLMs generating invalid variants (\textbf{C3}), we propose a dynamic-feedback-driven selection heuristic.
This heuristic integrates runtime status and dynamic analysis to filter out variants that fail to satisfy the constraints defined in the metamorphic relations.
% Furthermore, to overcome the limited variant generation capability of existing methods (\textbf{C3}),
% \sys{} leverages the extensive pre-training of LLMs on large-scale code corpora.
% This enables the generation of highly diverse program variants that naturally conform to the syntax and semantics of the target language.
% A key extension in \sys{} is its application of ideas from program repair and vulnerability injection to guide variant generation.
% This strategy is not confined to semantics-preserving transformations, which substantially increases the variety and fault-detection capability of the generated test cases.

We conducted a comprehensive evaluation of \sys{} on six SAST tools that support both Java and Rust programming languages. 
For each tool, we benchmarked our approach against existing testing techniques.
In total, \sys{} can detect \faultcnt{} unique faults in the tested SAST tools, with \fixedfaultcnt{} of these confirmed or fixed by the respective developers as previously unknown. 
The results demonstrate that \sys{} identifies the highest number of faults across both languages, discovering 3.4 times as many faults as the existing approach while requiring only 39.8\% of the variants.
Overall, our work makes the following main contributions:
\begin{itemize}
\item We propose \sys{}, to the best of our knowledge, the first LLM-driven automated testing framework for detecting faults in SAST tools.
It leverages the code knowledge encoded in large language models to generate diverse program variants, incorporating a dynamic feedback mechanism to validate their correctness.
Capitalizing on the multilingual nature of LLMs, \sys{} provides a universal testing solution applicable across multiple programming languages.
\item We introduce novel test oracles powered by LLMs, which support not only diverse semantics-preserving transformations but also program repair and vulnerability injection.
This approach significantly expands the semantic scope of generated variants beyond equivalence preservation, enabling the detection of deeper semantic faults.
The evaluation results demonstrate that \sys{} achieves 3.4 times higher fault detection effectiveness compared with the state-of-the-art technique.
\item We implemented \sys{} and conducted a large-scale evaluation on six mainstream open-source SAST tools for Java and Rust.
The framework successfully detected \faultcnt{} unique faults, of which \fixedfaultcnt{} have been confirmed or fixed by developers following our reports.
The results demonstrate the practical effectiveness and significant value of our approach. The experimental data and source code are available at: https://llm4sast.github.io/.
\end{itemize}

\section{Illustrative Example}
\label{sec:example}
We demonstrate the detailed workflow of \sys{} through a fault detected in \lockbud{}~\cite{lockbud2020pldi}, which is a popular SAST tool for Rust programming language.
Given a target SAST tool \textit{S} (e.g., \lockbud{}), \sys{} begins by extracting seed programs from the test suite or documentation of \textit{S}.
Figure~\ref{fig:example_left} illustrates a representative seed program taken from the official test suite.
The \lockbud{} developers designed program \textit{P} to test the ability of the tool to detect Use After Free (UAF) vulnerabilities.
In this program, the pointer \textit{raw\_ptr} referencing the local variable \textit{data} escapes the method \textit{func}, while \textit{data} is freed due to the ownership mechanism of Rust.
\lockbud{} correctly reports this security vulnerability, confirming the expected behavior of the tool for this test case.

% \begin{minipage}[t]{0.48\textwidth}
% \begin{figure}[H]
% \centering
% \begin{lstlisting}[language=rust]
% fn func() -> *const i32 {
%   let data=vec![1, 2, 3];
%   let ptr=data.as_ptr();
%   unsafe {
%     println!("{}", *ptr);
%   }
%   ptr
% }
% fn main() {
%   let ptr=func();
%   unsafe {
%     println!("{}", *ptr);
%   }
% }
% \end{lstlisting}
% \caption{A Rust UAF vulnerability example}
% \label{fig:image1}
% \end{figure}
% \end{minipage}
% \hfill
% \begin{minipage}[t]{0.48\textwidth}
% \begin{figure}[H]
% \centering
% \begin{lstlisting}[language=rust]
% use std::cell::RefCell;
% use std::rc::Rc;
% fn func() -> Rc<RefCell<Vec<i32>>> {
%   let data=Rc::new(
%     RefCell::new(vec![1, 2, 3])
%   );
%   // report a false warning
%   let ptr=data.as_ref().borrow().as_ptr();
%   unsafe {
%     println!("{}", *ptr);
%   }
%   data
% }
% fn main() {
%   let data=func();
%   println!("{:?}", data.borrow());
% }
% \end{lstlisting}
% \caption{Fixed UAF Code Example}
% \label{fig:uaf_example}
% \end{figure}
% \end{minipage}

\begin{figure}[htbp]
\centering
\begin{subfigure}[t]{0.46\linewidth}
\begin{minted}[fontsize=\small,numbersep=4pt,xleftmargin=1em]{rust}
fn func() -> *const i32 {
  let data=vec![1, 2, 3];
  let ptr=data.as_ptr();
  unsafe {
    println!("{}", *ptr);
  }
  ptr
}
fn main() {
  let ptr=func();
  unsafe {
    println!("{}", *ptr);
  }
}
\end{minted}
\caption{An Example of A Rust Use-After-Free Vulnerability}
\label{fig:example_left}
\end{subfigure}
\hfill
\begin{subfigure}[t]{0.46\linewidth}
\begin{minted}[fontsize=\small,numbersep=4pt,xleftmargin=1em]{rust}
use std::cell::RefCell;
use std::rc::Rc;
fn func() -> Rc<RefCell<Vec<i32>>> {
  let data=Rc::new(
    RefCell::new(vec![1, 2, 3])
  );
  // report a false warning
  let ptr=data.as_ref().borrow().as_ptr();
  unsafe {
    println!("{}", *ptr);
  }
  data
}
fn main() {
  let data=func();
  println!("{:?}", data.borrow());
}
\end{minted}
\caption{Fixed Use-After-Free Code Example}
\label{fig:example_right}
\end{subfigure}
\caption{An Illustrative Example of \sys{}}
\label{fig:uaf_pair}
\end{figure}

Existing approaches for testing SAST tools~\cite{cuoq2012testing, statfier, annatester, sascope, tdsc2024query, 2022icpcfinder, ase2024minotaur} fail to detect this category of faults because they lack the capability to repair the original vulnerability and depend on specialized oracles.
Specifically, random testing~\cite{cuoq2012testing} generates and selects mutants arbitrarily, which often results in inefficient exploration of variant spaces that are unlikely to reveal new faults in SAST tools.
Metamorphic testing techniques~\cite{statfier, annatester} are limited to semantics-preserving transformations or to faults triggered by specific language features; consequently, they cannot generate the fixed program variants required to expose such defects.
Differential testing~\cite{2022icpcfinder} relies on the existence of common detection rules across different SAST tools.
However, most rules are not fully equivalent and cannot be directly paired.
For instance, while \lockbud{} can directly detect use-after-free vulnerabilities, tools such as \semgrep{} and \prusti{} may lack this specific capability.
Furthermore, all these existing approaches are designed for specific programming languages, thus lacking support for a universal testing framework for SAST tools.

To address the limitations of prior approaches, we propose two key techniques: (1) context guided prompt synthesis and (2) dynamic-feedback-driven variant selection.
We elaborate on these techniques using the example presented in Listing~\ref{fig:uaf_pair}.

% \noindent \textbf{Context Guided Prompt Synthesis.}
\noindent \textbf{LLM-based Program Transformation with Diverse Semantics.}
% \yp{fixed: How about naming this contribution "LLM-based variant construction" or "LLM-based program transformation"? }
Prior research has demonstrated the strong capability of large language models in program transformation, fault repair, and vulnerability injection~\cite{foster2025mutation, bouzenia2024repairagent, dsn2024inject}.
To leverage these advancements, \sys{} utilizes LLMs to construct three types of metamorphic relations with corresponding program transformations and proposes an context guided approach for the automated synthesis of instructive prompts.
The key insight behind our prompt design is to help the LLM to quickly identify the buggy code snippets within the input program and to precisely understand the semantics of the vulnerability.

Usually, given an input program, a SAST tool generates an analysis report~$\mathcal{R}$, which contains details about the locations and types of potential security vulnerabilities and thereby satisfies the prerequisites for our approach.
Using program repair case~\ref{fig:example_left} as an example, given an input program \textit{P} containing a Use After Free (UAF) vulnerability, \sys{} first invokes the SAST tool \textit{S} to generate an analysis report $\mathcal{R}_S(P)$.
This report specifies the location information \textit{loc} where the UAF vulnerability occurs, i.e., lines 11--13 and columns 20--24 in Figure~\ref{fig:example_left}.
After obtaining the analysis report, \sys{} retrieves the documentation of the SAST tool to acquire a description of the UAF and systematically constructs the prompt based on the designed template for the program-repair metamorphic relation (see Figure~\ref{fig:mr2_template} for more details).

Subsequently, \sys{} submits the constructed prompt to the large language model to guide its repair of the UAF vulnerability in the program depicted in Figure~\ref{fig:example_left}.
The underlying rationale of this metamorphic relation is that, when the modified program is analyzed by the SAST tool, a successful repair should lead to the removal of the corresponding warning from its analysis report.
The program repaired by the LLM is presented in Figure~\ref{fig:example_right}.
The model resolves the issue by extending the lifetime of the local variable \textit{data} using smart pointers in Rust, namely \textit{Rc} and \textit{RefCell}.
This modification ensures that the variable \textit{data} returned by the function \textit{func} remains accessible within the \textit{main} function, thereby eliminating the UAF vulnerability.

\noindent \textbf{Dynamic-Feedback-Driven Variant Selection.}
Although the correctness of the repair in Figure~\ref{fig:example_right} can be manually verified through behavioral and semantic analysis, automated validation of fixed program variants remains challenging due to the absence of reliable oracles.
To overcome this limitation, we introduce our first heuristic: dynamic-feedback-driven variant selection.
The vulnerable program in Figure~\ref{fig:example_left} can exhibit undefined behavior when accessing a freed object.
To detect such issues, we utilize the dynamic analysis tool \miri{}~\cite{miri}, which generates an analysis report $\mathcal{R}_2$.
By comparing the location and vulnerability type information between reports $\mathcal{R}_1$ (from \lockbud{}) and $\mathcal{R}_2$ (from \miri{}), we verify whether both tools identify the same underlying defect.
Once the fix is confirmed, we analyze the repaired program (Figure~\ref{fig:example_right}) using \miri{}.  
The absence of the previously detected UAF vulnerability in the current analysis verifies the correctness of the fix.
Notably, \lockbud{} persists in flagging a UAF warning for the repaired program, which we classify as a false positive (FP) fault.
This discrepancy reveals a limitation in the analysis precision of \lockbud{}.

Furthermore, we have reported this fault to the development team of \lockbud{} via the GitHub issue tracking system~\cite{lockbud110}.
The developers confirmed this issue within one day and have added it to their fix schedule, attributing the root cause to breaking changes introduced by an update to the Rust toolchain, which violated certain invariants in the analysis tool.

\section{Methodology}
\label{sec:methodology}

% Figure~\ref{fig:workflow} provides an overview of \sys{} workflow.
% Given a SAST tool \textit{S} and its documentation and an input program, \sys{} first invokes \textit{S} to generate an analysis report of this input program, then it applies three types of semantics-diverse program transformations (semantics-preserving transformation, program-repair transformation, and vulnerability-injection transformation) to generate program variants.
% These three program transformations are relevant to defined metamorphic relations stated in Section~\ref{sec:mr}.
% The variant generation process utilizes the context guided prompt templates to synthesize prompts, which will be submitted to LLMs; an example is the program repair template illustrated in Figure~\ref{fig:pt2_template}.
% Next, for the variants generated by the LLM, \sys{} first checks each program for compilation errors, then uses an LLM to correct them.
% After that, \sys{} employs a dynamic-feedback-driven heuristic to filter out invalid instances.
% Owing to LLM hallucinations~\cite{xu2024hallucination}, such invalid variants often fail to satisfy the requirements specified in the prompts.
% Finally, since the LLM can produce a large number of variants, we propose a program-differencing-based clustering method to efficiently identify unique faults.

\begin{figure}[!htbp]
\centering
\includegraphics[width=\linewidth]{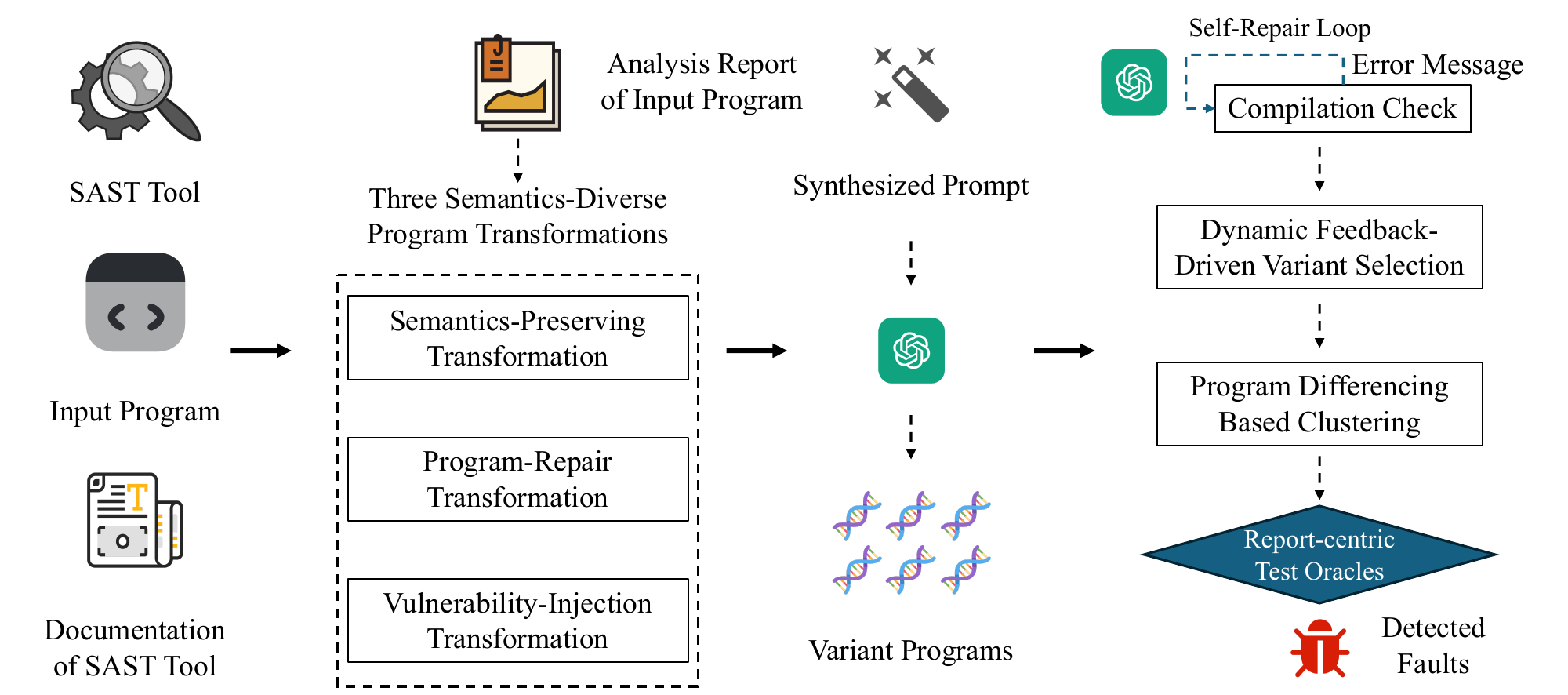}
\caption{Overview of \sys{} Workflow}
\label{fig:workflow}
\end{figure}

Figure~\ref{fig:workflow} presents an overview of the \sys{} workflow.
Given a SAST tool \textit{S} together with its documentation and an input program, \sys{} first invokes \textit{S} to generate an analysis report for the program.
It then applies three types of semantics-diverse program transformations (semantics-preserving transformation, program-repair transformation, and vulnerability-injection transformation) to generate program variants.
These transformations correspond to the metamorphic relations defined in Section~\ref{sec:mr}.
The variant generation process leverages context-guided prompt templates to synthesize prompts, which are submitted to LLMs; for example, Figure~\ref{fig:mr2_template} illustrates the program-repair template.
For each variant generated by the LLM, \sys{} first checks for compilation errors and uses the LLM to correct them when necessary.
Subsequently, \sys{} applies a dynamic-feedback-driven heuristic to filter out invalid instances.
Such invalid variants often arise from LLM hallucinations~\cite{xu2024hallucination} and fail to meet the requirements specified in the prompts.
Finally, since the LLM can generate a large number of variants, we introduce a program-differencing-based clustering method to efficiently identify unique faults.

\begin{algorithm}
\footnotesize
\SetKwInput{KwInput}{Input} % Set the Input
\SetKwInput{KwOutput}{Output} % Set the Output
\DontPrintSemicolon
\KwInput{Input programs $\mathit{Progs}$ for a rule $R$ in an SAST tool $S$, and the similarity threshold $\theta$}
\KwOutput{A set $errP$ with test programs indicating potential bugs in the tested SAST tool}
% Set Function Names。 
\SetKwFunction{FMain}{variantSelect}
\SetKwFunction{FLoc}{\textsc{isDifferential}}
\SetKwProg{Pn}{Func}{:}{\KwRet}
\SetKwProg{Fn}{Func}{:}{\KwRet}
\SetKw{Continue}{continue}
\SetKw{Break}{break}
$\mathit{Clusters} \leftarrow \emptyset$\\
$\mathit{errP} \leftarrow \emptyset$\\
$\mathit{Trans} \leftarrow [\mathit{Trans_1}, \mathit{Trans_2}, \mathit{Trans_3}] $\\
% $\mathit{Trans\_Set} \leftarrow [\Phi_{sp}, \Phi_{pr}, \Phi_{vi}] $\\
% \textsc{initializeTimer($\mathit{execTime}$)}\\
\For{$currP \in Progs$} {
    % $\mathit{Q}$ $\leftarrow$ \textsc{init\_queue($P$)}\\
    % \While{$\mathit{execTime} \leq timeL \land Q \neq \emptyset $} {
        % $\mathit{normP} \leftarrow \emptyset $\\
        % $\mathit{currP}$ $\leftarrow$ $Q$.\textsc{dequeue()}\\
    % ($\mathit{numW_{B}}, \mathit{typeW_{B}}, \mathit{loc_{B}}$) $\leftarrow \mathit{runSAST}$($\mathit{currP}$, $\mathit{S}$)\;
    $\mathcal{R} \leftarrow$ \textsc{runSAST}$(\mathit{currP}, \mathit{S})$\;
    \For{$\mathit{T} \in \mathit{Trans}$} {
        $\mathit{prompt} \leftarrow$ \textsc{PromptSynthesis}($\mathit{currP}, \mathcal{R}, \mathit{Trans}$)\;
        $\mathit{newProgs} \leftarrow$ \textsc{VariantGeneration}($\mathit{prompt}$)\;
        \For{$\mathit{newP} \in \mathit{newProgs}$} {
            \If{$\lnot$\textsc{isValidVariant{$(\mathit{currP}, \mathit{newP}$)}}} {
                \Continue\;
            }
            \If{\textsc{isFaultRevealing}($\mathit{T}, \mathit{newP}, \mathcal{R}$)}{
                $\mathit{errP} \leftarrow \mathit{errP} \cup \{\mathit{newP}\}$ \\
                $\mathit{fv}$ $\leftarrow$ \textsc{ProgramDiff($\mathit{currP}, \mathit{newP}$)} \\
                $(\mathit{\theta}, C) \leftarrow$ \textsc{CalculateMaxSimilarity($\mathit{fv}, \mathit{Clusters}$)} \\
                \If{$\mathit{MaxSim} > \mathit{\theta}$} {
                    $C \leftarrow C \cup \mathit{fv}$ \tcp{Assign the feature vector (fv) to an existing cluster (C)}
                } \Else {
                    $\mathit{Clusters} \leftarrow \mathit{Clusters} \cup \mathit{\{fv\}}$ \tcp{Set \textit{fv} as the centroid of a new cluster}
                }
            }
                % \Else {
                %     $\mathit{normP} \leftarrow \mathit{normP} \cup \mathit{newP}$
                %     % \tcc*{dynamic validation and store new variants}
                % }
        }
   }
       % $\mathit{Q}$ $\leftarrow$ $Q.\textsc{enqueue($\mathit{normP}$)}$
    % }
}
\KwRet $errP$\;
% \Pn{\textsc{IsFault} ($\mathit{newP}$, $\mathit{numWarn}_{B}$, $\mathit{typeWarn}_{B}$)} {
\Pn{\textsc{isFaultRevealing} ($\mathit{T}, \mathit{newP}$, $\mathcal{R}$)} {
    % $(\mathit{numWarn_{A}}, \mathit{typeWarn_{A}}, \mathit{loc_{A}}) \leftarrow$ \textsc{runSAST}$(\mathit{newP})$\\
    $\mathcal{R'} \leftarrow$ \textsc{runSAST}$(\mathit{newP}, \mathit{S})$\\
    \If {$\mathcal{R'}.\mathit{type} \setminus \mathcal{R}.\mathit{type} \neq \emptyset$} {
        \KwRet $False$
    }
    \Switch{$T$} {
        \Case{$\mathit{Trans_1}$} {
            \KwRet $\mathcal{R} = \mathcal{R'}$ \tcp{Metamorphic Relation 1}
        }
        \Case{$\mathit{Trans_2}$} {
            \KwRet $\mathcal{R'} \subset \mathcal{R}$ \tcp{Metamorphic Relation 2}
        }
        \Case{$\mathit{Trans_3}$} {
            \KwRet $\mathcal{R} \subset \mathcal{R'}$ \tcp{Metamorphic Relation 3}
        }
    }
    % \If{$T=\mathit{Trans_1}$} {
    %     \KwRet $\mathcal{R} = \mathcal{R'}$ \tcp{Metamorphic Relation 1}
    %     % \KwRet $\mathit{numWarn_{B}} \neq \mathit{numWarn_{A}} \lor \mathit{typeWarn_{B}} \neq \mathit{typeWarn_{A}}$
    % }
    % \If{$T=\mathit{Trans_2}$} {
    %     \KwRet $\mathcal{R'} \subset \mathcal{R}$ \tcp{Metamorphic Relation 2}
    %     % \KwRet $\mathit{numWarn_{B}} > \mathit{numWarn_{A}} \lor \mathit{typeWarn_{B}} > \mathit{typeWarn_{A}}$
    % }
    % \If{$T=\mathit{Trans_3}$} {
    %     \KwRet $\mathcal{R} \subset \mathcal{R'}$ \tcp{Metamorphic Relation 3}
    %     % \KwRet $\mathit{numWarn_{B}} < \mathit{numWarn_{A}} \lor \mathit{typeWarn_{B}} < \mathit{typeWarn_{A}}$
    % }
}
\caption{Algorithm for LLM-driven Automated Testing of SAST Tools}
\label{alg:workflow}
\end{algorithm}

% \yp{the algorithm involves too many details. How about moving it to the end of the section?}
% \zhe{I try to simplify this part and some details have been moved to the following subsections.}

% Algorithm~\ref{alg:workflow} details the implementation of \sys{}.
% Given a set of input programs $Progs$ and a SAST tool $S$, our testing algorithm consists of three core components: (1) LLM-based multi-language program transformation, (2) dynamic-feedback-driven variant selection, and (3) program-differencing-based clustering.
% The algorithm aims to generate variants that can reveal faults in $S$.

Algorithm~\ref{alg:workflow} presents the implementation of \sys{}. 
Given a set of input programs $Progs$ and a SAST tool $S$, the algorithm comprises three core components: 
(1) LLM-based multi-language program transformation engine, 
(2) dynamic-feedback-driven variant selection, and 
(3) program-differencing-based clustering. 
These components work together to produce program variants capable of uncovering faults in $S$.

% \sys{} first initializes the algorithm parameters (lines 1--3).
% Then, for each seed program \textit{P} from \textit{Progs}, \sys{} invokes the SAST tool $S$ to analyze it and obtains an analysis report.
% This report includes the types and detailed locations of detected vulnerability warnings (lines 4--5).
% \sys{} subsequently selects a program transformation (line 6) to determine the corresponding prompt template.
% It then leverages the vulnerability type information from the analysis report to retrieve relevant vulnerability descriptions from the official documentation of \textit{S}.
% This contextual information is combined to construct a prompt, which is submitted to the LLM for variant generation (lines 7--8).

\sys{} first initializes the algorithm parameters (lines~1--3). 
For each seed program \textit{P} in \textit{Progs}, \sys{} invokes the SAST tool $S$ to analyze the program and obtains an analysis report. 
The report contains the types and precise locations of detected vulnerability warnings (lines~4--5). 
Next, \sys{} selects an appropriate program transformation \textit{T} corresponding to a specific type of metamorphic relation (line~6) to determine the prompt template.
Using the vulnerability type information from the report, it retrieves relevant vulnerability descriptions from the official documentation of \textit{S}.
This contextual information is then combined to construct a prompt, which is submitted to the LLM to generate program variants (lines~7--8).

Afterwards, for each newly generated program variant $\textit{newP}$, \sys{} applies the dynamic-feedback-driven selection heuristic to filter out invalid variants and determines whether the remaining ones can trigger faults in $S$ (lines~9--13).  
Subsequently, at lines~14--19, \sys{} performs cluster analysis on the validated variants to identify unique faults, thereby reducing the effort required to analyze the evaluation results (see Section~\ref{sec:program_diff} for details).

The core logic of the oracle design is implemented in the function \textsc{IsFault} (lines~21--31).
\sys{} first invokes the SAST tool \textit{S} to analyze the program variant \textit{newP}, and filters out any variants that introduce new vulnerability types compared with the original seed program.
This step is based on the key insight that correct program transformations, including semantics-preserving changes, vulnerability repairs, or vulnerability injections of the same type, should not cause \textit{S} to report new vulnerability categories.
program variants that fail to satisfy this criterion are deemed erroneous outputs from the large language model.
Next, \sys{} applies the three defined metamorphic relations to verify the correctness of the generated analysis reports (lines~25--31).
Variants that violate these relations are flagged as potential faults, and the details of the metamorphic relations along with their expected outcomes are provided in Section~\ref{sec:mr}.
% \yp{Are lines 21 and 22 necessary?}

In the following subsections, we elaborate on the details of each technique employed in the workflow of \sys{}.
% \yp{fixed: Use a subsection to introduce the three metamorphic relations, including the program transformations and their expected impacts on the SAST reports.}

\subsection{Metamorphic Relations and Expected Results}
\label{sec:mr}
In this subsection, we illustrate the implementation details of three metamorphic relations and their expected outcomes.
First, we define the analysis report generated by a SAST tool for a given input program.
% , and introduce the concept of analysis equivalence between two programs.

\begin{definition}[Analysis Report]
\label{def:cn}
% The analysis report generated by a SAST tool contains various types of warnings.
Given an input program $P$ and a SAST tool $S$, the analysis report generated by $S$ on $P$, denoted as $\mathcal{R}_S(P)$, is formally a set of warnings, with each warning defined by its type and location.
In other words, $\mathcal{R}_S(P) = \{\langle type_1,loc_1 \rangle,\ldots,\langle type_n,loc_n \rangle\}$.
% \yp{If we always only consider the type information, maybe we can simply define an analysis report as a set of issue types. The reason should be explained.}
% $\mathcal{R}_S(P)$ composed by different pairs $\{\mathit{<type_i, loc_i>}\}$.
% Each pair corresponds to a unique warning, where $\mathit{type_i}$ and $\mathit{loc_i}$ denote the type and source code location of a warning, respectively.
\end{definition}

% \begin{definition}[Analysis Equivalence]
% \label{def:cn}
% Two programs $P$ and $P'$ are \emph{analysis equivalent} with respect to a SAST tool $S$, denoted as $P \equiv_S P'$, if and only if:
% \begin{enumerate}
%     \item $S$ reports the same set of issues for $P$ and $P'$, i.e., $\mathcal{R}_S(P) = \mathcal{R}_S(P')$ for all $i$; 
%     and
%     \item $S$ terminates in the same state (either successfully or with errors) when applied to $P$ and $P'$.
%     \yp{The other two MRs do not need to be concerned with the termination states?}
% \end{enumerate}
% We write $P \equiv P'$ to denote that $P$ and $P'$ are analysis equivalent with respect to any SAST tool.
% \end{definition}

% \begin{definition}[Analysis Equivalence]
% \label{def:cn}
% Two programs $P$ and $P'$ are analysis equivalent w.r.t.\ a SAST tool $S$, denoted as $P\equiv_S P'$, if and only if (1) $S$ reports the same issues on $P$, i.e., $\mathcal{R}_S(P).type_i=\mathcal{R}_S(P').type_i$ and $P'$ and (2) $S$ terminates in the same state, i.e., successful or with errors when applied on $P$ and $P'$.
% We use $P\equiv P'$ to denote that $P$ and $P'$ are analysis equivalent w.r.t.\ any SAST tool.
% \end{definition}

% In the rest of this subsection, we use the following notations.
% Let $P$ be a seed program, and $\psi(P)$ be the set of generated variants after applying program transformation $\psi$ to $P$.
% and we define the following three metamorphic relations between $P$ and a variant from 
% Overall, we define three types of metamorphic relations (MRs):

\begin{definition}[Metamorphic Relation (MR)]
\label{def:mrs}
% \mbox{}%\vspace{-0.5\baselineskip}
Given a seed program $P$, a program transformation $\psi$, and a SAST tool $S$, 
we define the following metamorphic relations for the analysis performed by $S$, where $\psi(P)$ denotes the set of all $P$-variants generated by applying $\psi$ to $P$,
and $P'\in \psi(P)$.
Besides, \textit{S} should terminate in the same state (either successfully or with errors) when applied to $P$ and $P'$ for all metamorphic relations.
\begin{enumerate}
\item[\textbf{MR1:}] $\textrm{isSemPreserving}(\psi)\implies \mathcal{R}_S(P) = \mathcal{R}_S(P')$;
% Given a program $P$ and a semantics-preserving transformation $\psi_{sp}$, $\forall P' \in \psi_{sp}(P)$ should be analysis equivalent to $P$, i.e., $P \equiv P'$.
\item[\textbf{MR2:}] $\textrm{isProgRepair}(\psi) \implies \mathcal{R}_S(P')\subseteq \mathcal{R}_S(P) $
% Given a program $P$ and a program-repair transformation $\psi_{pr}$, $ \forall P' \in \psi_{pr}(P), \mathcal{R}_S(P').\text{type}_i \subseteq \mathcal{R}_S(P).\text{type}_i$ for all $i$.
\item[\textbf{MR3:}] $\textrm{isVulInjection} (\psi)\implies \mathcal{R}_S(P)\subseteq \mathcal{R}_S(P')$
% Given a program $P$ and a vulnerability-injection transformation $\psi_{vi}$, $ \forall P' \in \psi_{vi}(P), \mathcal{R}_S(P).\text{type}_i \subseteq \mathcal{R}_S(P').\text{type}_i$ for all $i$.
\end{enumerate}
\end{definition}

% \begin{definition}[MR2]
% \label{def:mr2}
% Given a program $P$ and a program repair transformation $\psi_{pr}$, the program variant $P'$ should possess the property $\mathcal{R}_S(P').{type_i} \subseteq \mathcal{R}_S(P).{type_i}, \forall P' \in \psi_{pr}(P)$.
% \end{definition}

% \begin{definition}[MR3]
% \label{def:mr3}
% Given a program $P$ and a vulnerability injection transformation $\psi_{vi}$, the program variant $P'$ should possess the property $\mathcal{R}_S(P).{type_i} \subseteq \mathcal{R}_S(P').{type_i}, \forall P' \in \psi_{vi}(P)$.
% \end{definition}

Intuitively, metamorphic relation \textbf{MR1} stipulates that, if $P'$ is derived from $P$ via a semantic-preserving transformation, the SAST tool $S$ should detect the same set of issues on $P$ and $P'$ and terminate in the same state;
% a semantic preserving variant $P'$ should be analysis equivalent to the corresponding seed program $P$, with respect to any SAST tool.
metamorphic relation \textbf{MR2} stipulates that, if $P'$ is derived by repairing a vulnerability in $P$, the issues the SAST tool $S$ detects on $P'$ should be a subset of what $S$ detects on $P$;
metamorphic relation \textbf{MR3} stipulates that, if $P'$ is derived by injecting a vulnerability into $P$, the issues the SAST tool $S$ detects on $P'$ should be a superset of what $S$ detects on $P$.
\sys{} detects defects in SAST tools based on these three metamorphic relations, as shown in Algorithm 1 (lines 25--31).
In particular, each violation to the metamorphic relations indicates a potential fault in the corresponding SAST tool $S$.

% After defining the three MRs, we validate the generated variants at lines~23--28 in Algorithm~\ref{alg:workflow}.
% For \textbf{MR1} with the semantics-preserving transformation, the number and types of vulnerability warnings\yp{Definition 4.1 only refers to ``warnings''. We should make sure the terms are consistent.} in both analysis reports should be identical; any discrepancy suggests a potential fault at lines~23--24.
% For \textbf{MR2} with the program-repair transformation, the number and types of warnings in the program variant should be a subset of those in the seed program, as checked at lines~25--26.
% For \textbf{MR3} with the vulnerability-injection transformation, the number and types of warnings in the program variant should be a superset of those in the seed program, as checked at lines~27--28.

% After defining three MRs, we validate the generated variants at lines 23--28 in Algorithm~\ref{alg:workflow}.
% For $\mathit{MR_1}$ and the semantics-preserving transformation, the number and types of vulnerability warnings in both analysis reports should be identical; the discrepancy indicates a potential fault at lines 23--24.
% For $\mathit{MR_2}$ and the program repair transformation, the number and types of reported warnings in the program variant should be a subset of those in the seed program at lines 25--26.
% For $\mathit{MR_3}$ and the vulnerability injection transformation, the number and types of reported warnings in the program variant should be a superset of those in the seed program at lines 27--28.

\subsection{Harnessing Large Language Models to Generate Variants}
This subsection describes the prompt synthesis and variant generation process.  
Given an input program $P$, \sys{} produces variants by submitting carefully designed prompts to a large language model (LLM) (lines~7--8 in Algorithm~\ref{alg:workflow}).  
The prompts are created based on three distinct metamorphic relations and their associated program transformations.  
To ensure the comprehensiveness of our prompt design, we first conduct a thorough review of relevant literature~\cite{promptdesign1, promptdesign2, promptdesign3, mao2025prompts, wang2025mut4all, openaipe} and adhere to the following principles when constructing the prompt templates.

\noindent \textbf{Sufficient and Precise Contextual Information.}
To mitigate the adverse effects of hallucinations in LLM-generated outputs, we provide LLMs with comprehensive contextual information derived from analysis reports and documentation of SAST tools.  
The analysis reports help LLMs accurately identify vulnerabilities in input programs, and \sys{} includes a report parser that extracts the precise location and type of each vulnerability from SAST tool analysis reports.
Furthermore, to ensure accurate understanding of the semantic details associated with detected vulnerabilities, we use the extracted type information to retrieve detailed descriptions from the official SAST tool documentation.

\noindent \textbf{Few-Shot Learning.}  
To enable LLMs to accurately perform semantic-preserving, program-repair, and vulnerability-injection program transformations, we adopt a few-shot learning paradigm.  
This approach provides targeted demonstrations that allow LLMs to process semantic constraints and operational patterns of code with minimal examples, thereby reducing reliance on large labeled datasets, facilitating efficient knowledge transfer, and enabling rapid adaptation to previously unseen downstream tasks.

\noindent \textbf{Well-Structured and Instructive Prompt Template.}  
To enhance the effectiveness of LLMs, we follow the recommendation from OpenAI~\cite{openaipe} and design a well-structured prompt template that incorporates clear instructions regarding the role, directive, workflow, context, examples, output formats, and constraints.
The role and directive are specified in the system prompt, while the remaining components are contained in the user prompt.

Following the design principles outlined above, we present the system prompt in Figure~\ref{fig:sysprompt} and the user prompt templates for the semantics-preserving metamorphic relation in Figure~\ref{fig:mr1_template}, the program-repair metamorphic relation in Figure~\ref{fig:mr2_template}, and the vulnerability-injection metamorphic relation in Figure~\ref{fig:mr3_template}.  
An illustrative example of the program-repair metamorphic relation has been discussed earlier in Section~\ref{sec:example}

% system prompt
\begin{figure}[htbp]
\centering
\begin{minipage}{\textwidth}
\begin{systemprompt}
You are a testing expert, and your task is transforming input programs for testing static application security testing (SAST) tools and constructing program variants that can trigger faults.
\end{systemprompt}
\end{minipage}
\caption{Overview of the System Prompt}
\label{fig:sysprompt}
\end{figure}

% semantics-preserving prompt
\begin{figure}[!htbp]
\centering
\begin{minipage}{\textwidth}
\begin{userprompt}
\textbf{[Function Instructions]:}\\
\textbf{1.} Understand the \texttt{\{Vulnerability\_Type\}} vulnerability by cross-referencing its document: \texttt{\{Document\_of\_Vulnerability\_Type\}};\\
\textbf{2.} Leverage the analysis report to both locate the vulnerability and comprehend its root cause, use program slicing to identify the relevant program elements;\\
\textbf{3.} Perform varied semantics-preserving transformations on these elements to produce diverse variants. Each transformation must preserve both the original program behavior and the reported count of \texttt{{Vulnerability\_Type}} warnings in the analysis report.\\
\textbf{[Output Instructions]}\\
The output must be source code of transformed programs, not add comments and new line breaks\\
\textbf{[Input]}\\
Here are the contents of the analysis report: \texttt{\{Analysis\_Report\}} and input program \texttt{\{Input\_Program\}}.
\end{userprompt}
\end{minipage}
\caption{Prompt Template for Semantics-Preserving Related Metamorphic Relation}
\label{fig:mr1_template}
\end{figure}

% Program repair template
\begin{figure}[htbp]
\begin{minipage}[t]{\textwidth}
% \begin{userpromptwithtitle}
\begin{userprompt}
\textbf{[Function Instructions]:}\\
\textbf{1.} Understand the \texttt{\{Vulnerability\_Type\}} vulnerability by cross-referencing its document: \texttt{\{Document\_of\_Vulnerability\_Type\}};\\
\textbf{2.} Leverage the analysis report to both locate the vulnerability and comprehend its root cause;\\
\textbf{3.} Fix it using diverse approaches, and provide the fixed programs, ensuring diversity, and use the provided reference pair \texttt{\{Buggy\_Program\}} and \texttt{\{Fixed\_Program\}} to confirm your understanding of the pattern and a standard fix method for the vulnerability.\\
\textbf{[Output Instructions]}\\
The output must be source code of transformed programs, not add comments and new line breaks\\
\textbf{[Input]}\\
Here are the contents of the analysis report: \texttt{\{Analysis\_Report\}} and input program \texttt{\{Input\_Program\}}.\\
\end{userprompt}
% \subcaption{}
\label{fig:mr2_inner_template}
\end{minipage}
\caption{Prompt Template for Program-Repair Related Metamorphic Relation}
\label{fig:mr2_template}
\end{figure}

\begin{figure}[!htbp]
\centering

\begin{minipage}{\textwidth}
\begin{userprompt}
\textbf{[Function Instructions]:}\\
\textbf{1.} Understand the \texttt{\{Vulnerability\_Type\}} vulnerability by cross-referencing its document: \texttt{\{Document\_of\_Vulnerability\_Type\}};\\
\textbf{2.} Leverage the analysis report to both locate the vulnerability and comprehend its root cause;\\
\textbf{3.} Inject a new instance of this vulnerability type through diverse implementation approaches, while ensuring variation across the generated programs;\\
\textbf{4.} Learn the injection process by comparing against the provided reference pair: \texttt{\{Buggy\_Program\}} and \texttt{\{Injected\_Program\}}.\\
\textbf{[Output Instructions]}\\
The output must be source code of transformed programs, not add comments and new line breaks\\
\textbf{[Input]}\\
Here are the contents of the analysis report: \texttt{\{Analysis\_Report\}} and input program \texttt{\{Input\_Program\}}.\\
\end{userprompt}
\end{minipage}

\caption{Prompt Template for Vulnerability-Injection Related Metamorphic Relation}
\label{fig:mr3_template}
\end{figure}

\subsection{Validation of Generated Program Variants}
\label{sec:post_validation}
Given the inevitable hallucination limitations of large language models~\cite{xu2024hallucination}, we implement a validation mechanism for generated variants to enhance the correctness of testing process.
Based on our observations, we categorize invalid program variants into two types.  
The detailed variant validation algorithm is presented in Algorithm~\ref{alg:validation}.

\begin{algorithm}
\footnotesize
\SetKwInput{KwInput}{Input} % Set the Input
\SetKwInput{KwOutput}{Output} % Set the Output
\DontPrintSemicolon
\KwInput{An input program $\mathit{currP}$, its variant program $\mathit{newP}$, the program transformation $\mathit{T}$ currently used.}
\KwOutput{A boolean value indicating whether the variant program is valid.}
% Set Function Names
\SetKwFunction{FMain}{variantSelect}
\SetKwFunction{FLoc}{\textsc{isDifferential}}
\SetKwProg{Pn}{Func}{:}{\KwRet}
\SetKwProg{Fn}{Func}{:}{\KwRet}
\SetKw{Continue}{continue}
\SetKw{Break}{break}
\Pn{\textsc{VariantValidation} ($\mathit{newP}$, $\mathit{currP}$, $T$)} {
$c \leftarrow $\textsc{compile}($\mathit{newP}$)\\
\If {$\neg c$} {
    \For {$i = 1 \rightarrow{} \mathit{fixL}$} {
        $\mathit{newP} \leftarrow $ \textsc{FixCompilationError}$(\mathit{newP})$\\
        \If {$\neg$\textsc{compile}($\mathit{newP}$)} {
            $c \leftarrow True$\\
            \Break\;
        }
    }
}
\If {$\neg c$} {
    \KwRet $False$\;
}
$\tau_1 \leftarrow{}$ \textsc{runProgram}($\mathit{currP}$)\\
$\tau_2 \leftarrow{}$ \textsc{runProgram}($\mathit{newP}$)\\
$\theta_1 \leftarrow{}$ \textsc{runDynamicAnalyzer}($\mathit{currP}$)\\
$\theta_2 \leftarrow{}$ \textsc{runDynamicAnalyzer}($\mathit{newP}$)\\
\If {$\mathit{T} = \mathit{Trans_1}$} {
    \If {$\tau_1 = \tau_2 \land \theta_1 = \theta_2$} {
        \KwRet $True$\;
    }
}
\If{$\mathit{T} = \mathit{Trans_2}$} {
    \tcc{A runtime status equal to 0 means a normal exit.}
    \If {$\tau_1 \neq 0 \land \tau_2 = 0 \land |\theta_1| \geq |\theta_2|$} {
        \KwRet $True$\;
    }
}
\If{$\mathit{T} = \mathit{Trans_3}$} {
    \If {$\tau_1 = \tau_2 \land |\theta_1| \leq |\theta_2|$} {
        \KwRet $True$\;
    }
}
\KwRet $False$;
}
\caption{Algorithm for Variant Validation}
\label{alg:validation}
\end{algorithm}

The first category (\textbf{C1}) comprises uncompilable program variants. 
As LLM-generated variants may fail to compile, we address \textbf{C1} by combining each uncompilable variant with its corresponding compilation error message to form a new prompt illustrated in Figure~\ref{fig:self_repair}.
This prompt is submitted to LLMs to rectify the errors.
The second category (\textbf{C2}) includes variants that fail to satisfy the semantic constraints specified in the prompts.
For example, in the context of the program-repair transformation, valid variants should not contain the vulnerability indicated in the prompt.  
To address \textbf{C2}, we identify three types of semantic constraints implied by the prompts and employ a dynamic-feedback-driven heuristic to verify compliance.
The dynamic feedback for a program $p$ consists of its runtime status $r$, determined by the return status code and execution time, and the results $d$ obtained from dynamic analysis tools (\miri{} for Rust and \texttt{jstack} for Java).
In the vulnerability-repair transformation, the $r$ and $d$ values of program variants should differ from those of the seed program if the seed program exhibits abnormal behavior.
Specifically, vulnerabilities in the seed program may lead to runtime exceptions or timeouts caused by abnormal memory access or deadlocks, whereas the status of the program variant should indicate that these issues have been successfully corrected.
Conversely, for the semantics-preserving and vulnerability-injection program transformations, the $r$ and $d$ of seed and program variants should remain consistent.
Consequently, we filter out program variants that do not satisfy these semantic constraints.

\begin{figure}[!htbp]
\centering
\begin{minipage}{\textwidth}
\begin{userprompt}
\textbf{[Function Instructions]}\\
\textbf{1.} You provide me a program variant previously: \texttt{\{Generated\_Code\}}\\
\textbf{2.} However, it has the following compilation errors: \texttt{\{Compilation\_Error\}}\\
\textbf{3.} Please fix them and give me a correctly compilable program.\\
\end{userprompt}
% \subcaption{}
\label{fig:userprompt}
\end{minipage}
\caption{Prompt Template for Self-Repair Loop}
\label{fig:self_repair}
\end{figure}

\subsection{Program-Differencing-Based Unique Fault Clustering}
\label{sec:program_diff}
During variant generation, large language models (LLMs) may apply identical transformations to similar program elements, resulting in multiple variants triggering the same fault.  
We address this issue by proposing a program-differencing-based unique fault clustering technique.  
For enhanced diversity in generated variants, we utilize an LLM to directly generate their derivatives, rather than relying on a fixed set of program transformations.  
While this approach leverages the extensive pre-trained code data available to the LLM, it complicates capturing the relationship between generated variants and their corresponding seed programs, specifically the type of program transformation applied.  
Consequently, we employ a program-differencing technique to identify their correlations, as detailed at lines 12--17 of Algorithm~\ref{alg:workflow}.

% First, it uses a program-differencing-based technique to extract the sequence of edit scripts between the seed program $\textit{currP}$ and the program variant $\textit{newP}$, and then leverages a pre-trained model to convert this sequence into a feature vector $\textit{fv}$.
% Next, the maximum similarity value $\theta$ between $\textit{fv}$ and the centroid vectors of existing clusters is calculated.
% If $\theta$ exceeds a threshold $T$, $\textit{fv}$ is merged into the corresponding cluster.
% If $\theta$ is less than $T$, $\textit{fv}$ is established as the centroid of a new cluster and is checked by the test oracles (lines 16--17).
% Finally, all program variants that reveal potential faults are returned (line 18).

At line 12 of Algorithm~\ref{alg:workflow}, for an input program \textit{currP} and its program variant \textit{newP}, we first apply the GumTree technique~\cite{gumtree} to calculate the edit scripts that capture code changes from \textit{currP} to \textit{newP}.  
Directly using raw edit scripts from GumTree for exact matching produces overly sparse features, as minor differences between variants hinder effective clustering.  
To address this, we abstract the edit scripts while preserving core syntactic and semantic information.
Initially, we normalize the edit scripts to reduce semantic noise and mitigate the curse of dimensionality.
Specifically, we replace detailed identifier names with placeholders, such as $VAR$ for variable names and \textit{LITERAL} for string or integer literals.
Subsequently, we consolidate syntactically similar abstract syntax tree (AST) node types to reduce granularity and improve generalizability.
For example, we unify \textit{TernaryExpression} and \textit{IfStatement}, both representing conditional logic, into a single \textit{Conditional} type.
Additionally, we merge consecutive edit script segments in a semantics-preserving manner to enhance abstraction.
For instance, combining a \textit{Delete} action followed by an \textit{Insert} action for the same object yields a unified \textit{Relocate} action.
This abstraction reduces the sparsity of change vectors and aligns the representation with developer intent.

After normalization, we employ QWen3-Embedding-0.6B~\cite{qwen3embedding, qwen3} to generate vector representations (embeddings) $\mathit{fv}$ of the normalized edit-script sequences.
These embeddings serve as quantitative representations of program transformations for clustering and are also mapped to each generated variant program.
Our motivation for choosing this model for vectorization stems from its performance in the Massive Text Embedding Benchmark (MTEB)~\cite{mteb} and the computational limitations of our local server.

The clustering process entails a linear scan of these feature vectors. 
For each vector $\mathit{fv}$, the maximum cosine similarity $\mathit{MaxSim}$ is computed between $\mathit{fv}$ and the centroid vectors of the existing clusters ($\mathit{Clusters}$ at line 14).
If the maximum similarity $\mathit{MaxSim}$ exceeds a predefined threshold $\theta$, 
we assign $\mathit{fv}$ to the corresponding cluster $C$, as detailed at lines 16--17 of Algorithm~\ref{alg:workflow}.
% \yp{maximum? how many similarity values are computed for each fv? Exceeds? it's $\theta<T$ in the algorithm.}
% \yp{what does |fv| in the algorithm mean?}
Otherwise, we designate \textit{fv} as the centroid of a new cluster at lines 18--19.
% \yp{is this step reflected in the algorithm?}
After the clustering stage, we randomly select one sample from each cluster as the representative fault.
This representative is subsequently subjected to manual analysis to determine whether it corresponds to a real fault detected by the SAST tools or a false positive generated by \sys{}, and to identify its root cause.
We first conduct a manual analysis to examine whether the generated variants satisfy the constraints specified in the prompt. 
For example, LLMs may fail to fix bugs in the input programs, causing \sys{} to report a spurious fault. 
Subsequently, we categorize the faults identified in the selected samples into three types: false positives, false negatives, and runtime crashes.
% Following the clustering stage, one sample from each cluster is randomly selected as the representative fault.
% This representative is then manually analyzed to determine whether it reveals a fault in the SAST tools or a false positive produced by \sys{}, and to identify its root cause.

% For instance, specific operator nodes within a \textit{BinaryExpression} (such as +, -, *, and /) are unified under a general \textit{BinaryExpr} type.
% This abstraction helps to focus on the overarching program structure rather than superficial syntactic variations.
% We further abstract part of fine-grained edit operations into higher-level semantic actions.
% This involves merging sequences of primitive operations that collectively represent a single logical change.

\section{Evaluation}
We applied \sys{} to six static application security testing (SAST) tools, namely \lockbud{}~\cite{lockbud2020pldi}, \semgrep{}~\cite{semgrep}, \prusti{}~\cite{prusti}, \sonar{}~\cite{sonarqube}, \infer{}~\cite{infer}, and \sbugs{}~\cite{spotbugs}, for investigating the following research questions (RQs):
\begin{itemize}[leftmargin=*]
    \item RQ1: How many unique faults can \sys{} find? 
    \item RQ2: Can the proposed heuristics in \sys{} prove effective in reducing the number of variant programs and manual analysis efforts?
    \item RQ3: How does \sys{} perform compared to existing research work?
\end{itemize}

\subsection{Dataset and Experimental Setup}
\label{sec:dataset}
We implemented \sys{} in over 4,000 lines of Python code.
% The system comprises four key components: prompt synthesizer, variant selection.
The system was developed using the LangChain version 0.3.25.
Considering the popularity, performance, and representativeness, we adopt GPT-4o as the base model for evaluation.
We collect input programs from the official test suites and documentation of each SAST tool to form the initial seed corpus.
% For each input program that needs execution, we set the time limit $timeL$ to be 15 minutes, the program exceeding the $timeL$.
Informed by the related work~\cite{ccs2014malware, ccs2018invetter}, we assign a value of 0.7 to the similarity threshold $\theta$ in Algorithm~\ref{alg:workflow}.
For each rule of every evaluated SAST tool, we use the configuration from the test case where available, falling back to its default configuration otherwise.
All experiments were conducted on a server equipped with an Intel(R) Xeon(R) CPU Max 9468, 2 TB of RAM, and eight NVIDIA H100 GPUs.
Besides, our study focuses exclusively on security-related rules; consequently, non-security rules provided by the SAST tools are excluded from the testing process.
Table~\ref{tab:seedinfo} presents the statistical information of the collected seed programs in terms of the number of triggered vulnerability types and warnings.

\begin{table}[!htbp]
\caption{Statistics of Triggered Vulnerability Types and Warnings in Collected Seed Programs}
\label{tab:seedinfo}
\centering
% \setlength{\tabcolsep}{1pt}
% \resizebox{\linewidth}{!}{
\begin{tabular}{l|r|r|r|r|r|r|r}
\toprule
\textbf{Metrics} & \lockbud{} & \semgrep{} & \prusti{} & \sbugs{} & \infer{} & \sonar{} & \textbf{Total} \\
\midrule
\#Warnings & 48 & 291 & 113 & 635 & 358 & 124 & 1569 \\
\#Vulnerability Types & 6 & 57 & 4 & 46 & 19 & 18 & 150 \\
\bottomrule
\end{tabular}
% }
\end{table}

\noindent \textbf{Selection Criteria of Target SAST Tools.} We select SAST tools based on two criteria:
(1) it should be popular and widely adopted, ensuring their issues reflect real-world challenges faced by analyzer users.
Specifically, we focused on the top three SAST tools for both Java and Rust, as ranked by the number of GitHub stars;
(2) it must be open-source and maintain an issue tracking system to enable analysis of identified issues and communication with developers.
Based on these criteria, we choose Semgrep~\cite{semgrep}, LockBud~\cite{lockbud2020pldi}, and Prusti~\cite{prusti} for Rust analysis, along with Infer~\cite{infer}, SonarQube~\cite{sonarqube}, and SpotBugs~\cite{spotbugs} for Java.
Prusti~\cite{prusti} is a static verifier for Rust, based on the Viper verification infrastructure.
By default Prusti verifies absence of integer overflows and panics, proving that statements such as \textit{unreachable!()} and \textit{panic!()} are unreachable.
Semgrep~\cite{semgrep} is a fast, open-source, SAST tool that searches code, finds bugs, and enforces secure guardrails and coding standards.
Semgrep can run in an IDE, as a pre-commit check, and as part of CI/CD workflows.
Lockbud~\cite{lockbud2020pldi} is a static analysis tool designed to detect a variety of concurrency and memory safety issues in Rust code, including double-locks, conflicting lock orders, atomicity violations, and use-after-free.
Infer~\cite{infer} is an SAST tool developed by Meta for detecting potential bugs in Java or C/C++ programs, and its core technique roots in academic research on separation logic.
SonarQube~\cite{sonarqube} is an open-source platform developed by Sonar to integrate into software development workflows, ensuring continuous code quality and code security.
It provides continuous inspection of code quality to perform automatic reviews with static analysis of code to detect bugs, vulnerabilities, security hotspots, and code smells.
SpotBugs~\cite{spotbugs} is a program which uses static analysis to look for bugs in Java code, and it is the spiritual successor of FindBugs.
Although \sys{} can theoretically be applied to SAST tools supporting various programming languages theoretically, we focus on Java and Rust related tools as Java is the most widely used language, and Rust represents cutting-edge programming paradigms.

\subsection{RQ1: Assessing the Fault Detection Capability}
Fault detection capability is the most fundamental metrics for evaluating the performance of a testing approach.
In the first RQ, we present the fault-detecting capability of \sys{}, followed by root cause analysis of detected faults.
Overall, we have reported \faultcnt{} unique faults to developers of SAST tools.
Notably, we consider a fault to be a unique fault if it is in (1) different rule checkers triggered by various transformations (i.e., in different clusters stated in the section~\ref{sec:program_diff}),
(2) different faulty locations (determined by root cause diagnosis) in a SAST tool.
% and (3) not the \fpcnt{} listed in Section~\ref{}.
We use this definition based on the related work~\cite{statfier, fse2024sabugs} as well as developers' approaches to determining duplicate faults~\cite{sonarqubeguideline, pmdguideline}.
Table~\ref{tab:faultdetection} shows the current status of our submitted faults.
We classify these issues into three categories, namely ``Fixed'', ``Confirmed'', and ``Pending'', based on the responses that we have received from developers of SAST tools so far.

\begin{table}[!htbp]
\caption{The Status of Submitted Faults via GitHub}
\label{tab:faultdetection}
\centering
% \setlength{\tabcolsep}{1pt}
% \resizebox{\linewidth}{!}{
\begin{tabular}{l|r|r|r|r|r|r|r}
\toprule
\textbf{Status} & \lockbud{} & \semgrep{} & \prusti{} & \sbugs{} & \infer{} & \sonar{} & \textbf{Total} \\
\midrule
Fixed & 9 & 1 & 0 & 1 & 0 & 1 & 12 \\
Confirmed & 4 & 10 & 4 & 1 & 3 & 2 & 24 \\
Pending & 5 & 3 & 6 & 8 & 5 & 3 & 30 \\
\hline
\textbf{Total} & 18 & 14 & 10 & 10 & 8 & 6 & \faultcnt{} \\
\bottomrule
\end{tabular}
% }
\end{table}

\noindent\textbf{Distribution of Transformation Types.}
Furthermore, we evaluate the effectiveness of each type of semantics-diverse program transformation.
Table~\ref{tab:transformation} shows the distribution of detected faults across each transformation category for the six SAST tools.
The experimental results confirm that all three types of program transformations can detect unique faults in both programming languages.

\begin{table}[!htbp]
\caption{The Effectiveness of Each Category of Program Transformations}
\label{tab:transformation}
\centering
% \setlength{\tabcolsep}{1pt}
% \resizebox{\linewidth}{!}{
\begin{tabular}{l|r|r|r|r|r|r|r}
\toprule
\textbf{Transformation} & \lockbud{} & \semgrep{} & \prusti{} & \sbugs{} & \infer{} & \sonar{} & \textbf{Total} \\
\midrule
Semantics-preserving Transformation & 6 & 7 & 3 & 7 & 5 & 3 & 31 \\
Program-Repair Transformation & 7 & 4 & 5 & 2 & 1 & 0 & 19 \\
Vulnerability-Injection Transformation & 5 & 3 & 2 & 1 & 2 & 3 & 16 \\
\hline
\textbf{Total} & 18 & 14 & 10 & 10 & 8 & 6 & \faultcnt{} \\
\bottomrule
\end{tabular}
% }
\end{table}

\noindent\textbf{Symptoms of Detected Faults.}
To understand how these detected faults affect SAST tools, we analyze the symptoms of each fault.
Table~\ref{tab:symptom} shows the symptom distribution among six SAST tools.
All faults can be summarized into the following three categories.
\begin{itemize}
    \item \textbf{False Positive (FP)}: SAST tools provide the analysis reports with undesirable warnings;
    \item \textbf{False Negative (FN)}: SAST tools provide the analysis reports that are missing warnings;
    \item \textbf{Runtime Crash}: SAST tools prematurely terminate during execution.
\end{itemize}
The evaluation results shows that \sys{} not only can identify the runtime crash faults mainly due to front-end flaws, but also recognize faults that are related to the incorrect logic implementations, causing FP and FN faults.

\begin{table}[!htbp]
\caption{Symptom Distribution of Detected Faults Among Six SAST Tools}
\label{tab:symptom}
\centering
% \setlength{\tabcolsep}{1pt}
% \resizebox{\linewidth}{!}{
\begin{tabular}{l|r|r|r|r|r|r|r}
\toprule
\textbf{Symptom} & \lockbud{} & \semgrep{} & \prusti{} & \sbugs{} & \infer{} & \sonar{} & \textbf{Total} \\
\midrule
False Positive & 9 & 4 & 0 & 0 & 3 & 1 & 17 \\
False Negative & 9 & 7 & 0 & 10 & 5 & 4 & 35 \\
Runtime Crash & 0 & 3 & 10 & 0 & 0 & 1 & 14 \\
\bottomrule
\end{tabular}
% }
\end{table}

% \noindent \textbf{Fault-Triggering Characteristics in }
% We manually analyzed each minimized fault-triggering program to figure out what unique code features are in the real-world functions.

% \noindent\textbf{Variant Generation Capability.}
% In this part, we compare the variant generation capability of three categories of proposed program transformation.
% Table~\ref{} shows the detailed evaluation results.

\subsection{RQ2: Assessing the Effectiveness of Proposed Heuristics}
In this section, we try to assess the effectiveness of proposed two heuristics, namely dynamic-feedback-driven variant selection and program-differencing-based clustering.

The performance of the first heuristic is evaluated through an ablation study, using the number of detected faults, false positives, and variant programs as the evaluation metrics.
Table~\ref{tab:ablation-study} shows the experimental results of ablation study to compare the performance between original and ablated \sys{} versions.
As shown in this table, the dynamic-feedback-driven variant selection heuristic can help us reduce 47.8\% (12 v.s. 23) false positives while preserve the fault finding capability of \sys{} (both approaches identify \faultcnt{} faults), which enhance the performance and precision of \sys{}.
Besides, this heuristic can also reduce 54.5\% variants for further analysis, significantly improving the performance and usability of our tool.

\begin{table}[!htbp]
\caption{Ablation Study of Dynamic-Feedback-Driven Variant Selection Heuristic}
\label{tab:ablation-study}
\centering
% \setlength{\tabcolsep}{1pt}
% \resizebox{\linewidth}{!}{
\begin{threeparttable}
\begin{tabular}{l|l|r|r|r|r|r|r|r}
\toprule
\textbf{Metrics} & \textbf{Config} & \lockbud{} & \semgrep{} & \prusti{} & \sbugs{} & \infer{} & \sonar{} & \textbf{Total} \\
\midrule
\multirow{2}{*}{{\#}Unique Fault} & Original & 18 & 14 & 10 & 10 & 8 & 6 & \faultcnt{} \\ \cline{2-9}
& Ablated & 18 & 14 & 10 & 10 & 8 & 6 & \faultcnt{} \\
\hline
\multirow{2}{*}{{\#}False Positive} & Original & 3 & 3 & 0 & 3 & 1 & 2 & 12 \\ \cline{2-9}
& Ablated & 5 & 6 & 0 & 6 & 1 & 5 & 23 \\
\hline
\multirow{2}{*}{{\#}Variant} & Original & 498 & 1863 & 830 & 627 & 674 & 338 & 4830 \\ \cline{2-9}
& Ablated & 1067 & 3854 & 2070 & 1346 & 1737 & 548 & 10622 \\
\bottomrule
\end{tabular}
\begin{tablenotes}
\setlength{\itemindent}{0pt}
\footnotesize
\item In the second column, the original version indicates that the evaluation is performed using the dynamic-feedback-driven heuristic, while the ablated version denotes the opposite case (i.e., without this heuristic).
\end{tablenotes}
\end{threeparttable}
% }
\end{table}

For the second heuristic, we compare the number of generated variants and corresponding clusters to demonstrate its effectiveness.
Table~\ref{tab:cluster} shows the results of the evaluation.
According to the data in this table, the total number of generated variants across six SAST tools is 4830, and the total number of cluster is \clustercnt{}.
Consequently, our approach can reduce the effort required to review variants by approximately 98\% and significantly accelerate the overall testing process.
In addition, for each cluster, we analyzed all variant programs and found that their root causes matched those of the centering sample, confirming the effectiveness of this heuristic.

\begin{table}[!htbp]
\caption{Evaluation Data of Program-Differencing-Based Clustering}
\label{tab:cluster}
\centering
\resizebox{\linewidth}{!}{
\begin{threeparttable}
\begin{tabular}{l|cc|cc|cc|cc|cc|cc|cc}
\toprule
\multicolumn{1}{c|}{\multirow{2}{*}{Type}} & \multicolumn{2}{c|}{Lockbud} & \multicolumn{2}{c|}{Semgrep} & \multicolumn{2}{c|}{Prusti} & \multicolumn{2}{c|}{SpotBugs} & \multicolumn{2}{c|}{Infer} & \multicolumn{2}{c|}{SonarQube} & \multicolumn{2}{c}{\textbf{Total}} \\ \cline{2-15}
\multicolumn{1}{c|}{} & \#Variant & \#Cluster & \#Variant & \#Cluster & \#Variant & \#Cluster & \#Variant & \#Cluster & \#Variant & \#Cluster & \#Variant & \#Cluster & \#Variant & \#Cluster \\
\midrule
T1 & 232 & 10 & 702 & 10 & 256 & 6 & 160 & 11 & 307 & 7 & 174 & 8 & 1831 & 52 \\
T2 & 129 & 7 & 536 & 6 & 280 & 5 & 141 & 4 & 263 & 4 & 78 & 0 & 1427 & 26 \\
T3 & 137 & 8 & 625 & 4 & 294 & 3 & 326 & 2 & 104 & 2 & 86 & 5 & 1572 & 24 \\
\hline
\textbf{Total} & 498 & 25 & 1863 & 20  & 830 & 14 & 627 & 17 & 674 & 13 & 338 & 13 & 4830 & \clustercnt{} \\
\bottomrule
\end{tabular}
\begin{tablenotes}
\setlength{\itemindent}{0pt}
\item T1, T2, and T3 represent semantics-preserving, program repair, and vulnerability injection transformations.
\end{tablenotes}
\end{threeparttable}
}
\end{table}

\subsection{RQ3: Comparison with Existing Research Work}
% In this RQ, we try to compare \sys{} with existing researches for testing SAST tools.
% Among related researches~\cite{statfier, tdsc2024query, ase2024minotaur, ase2024sherlock, 2022icpcfinder, sascope}, we select Statfier~\cite{statfier} and BugFinder~\cite{2022icpcfinder} as they are the representative work of metamorphic and differential testing and purely focus on fault detection in SAST tools.
In this section, we compared \sys{} against Statfier~\cite{statfier}, which is the most recent and powerful testing framework for SAST tools.
Since this tool originally supported only the Java programming language, we adapted their core concepts and implemented a proof-of-concept version Statfier-Rust, which is capable of analyzing and transforming Rust programs.
Overall, Statfier-Rust supports 17 types of program transformations, comprising 10 directly adapted from Statfier and 7 designed by considering both the design principles of Statfier and the distinctive features of the Rust language.
For all three evaluated approaches, we use seed programs that were collected from the official test suites and documentation, as introduced in Section~\ref{sec:dataset}.

\begin{figure}[htbp]
    \centering
    \begin{subfigure}[t]{0.45\textwidth}
        \centering
        \includegraphics[width=\linewidth]{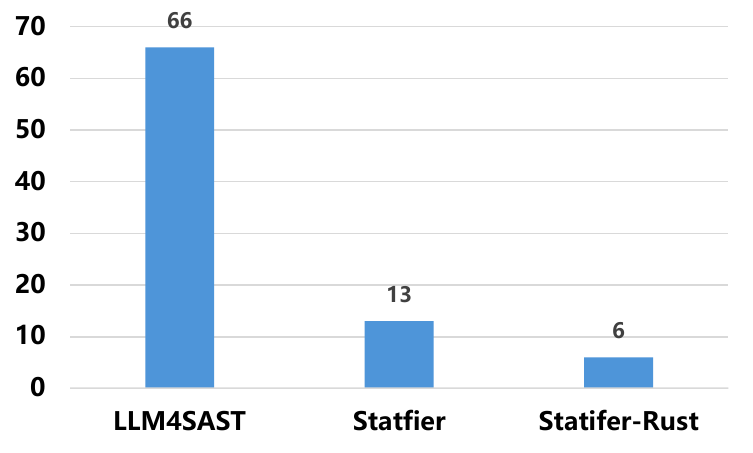}
        \caption{Number of Detected Faults for Each Approach}
        \label{fig:cnt_fault}
    \end{subfigure}
    \hfill
    \begin{subfigure}[t]{0.45\textwidth}
        \centering
        \includegraphics[width=\linewidth]{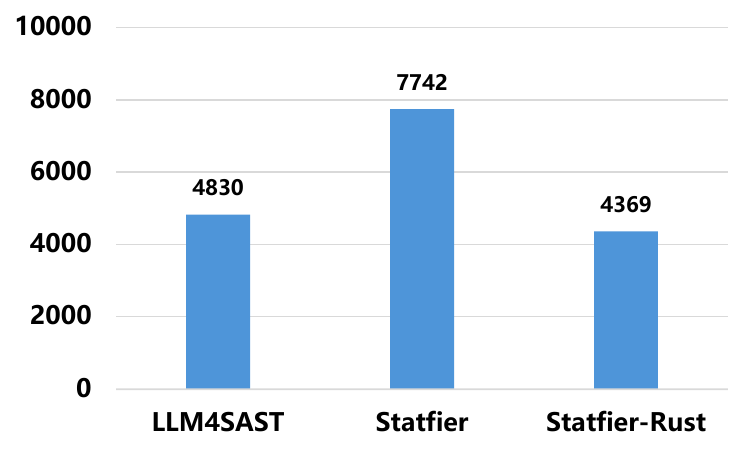}
        \caption{Number of Generated Variants for Each Approach}
        \label{fig:cnt_variant}
    \end{subfigure}
    \caption{Comparison Evaluation Among \sys{}, Statfier, and Statfier-Rust}
    \label{fig:comparison} 
\end{figure}

Figure~\ref{fig:comparison} presents the overall detection results of the three testing tools.
The left subfigure~\ref{fig:cnt_fault} shows the number of faults detected by each approach.
Our system, \sys{}, detects 5 times more faults than Statfier and over 10 times more than Statfier-Rust, confirming the effectiveness of our method.
It is worth noting that Statfier supports only semantic-preserving transformations.
In contrast, LLM4SAST additionally supports two types of semantic-altering transformations (program repair and vulnerability injection transformations).
Issues detected through semantic-altering transformations are typically difficult to uncover using purely semantic-preserving approaches, highlighting the capability of \sys{} in revealing faults that may otherwise remain hidden.

The right subfigure~\ref{fig:cnt_variant} displays the number of variants generated by each approach.
As shown, \sys{} requires only 73 variants per detected fault on average (4830/66), compared to 596 for Statfier (7742/13) and 728 for Statfier-Rust (4369/6).
This indicates that \sys{} achieves approximately 8.1$\times$ and 9.8$\times$ higher fault-finding efficiency than the SOTA approach Statfier and Statfier-Rust, respectively.

\section{Discussion}

\subsection{Extension of \sysb{}}
Our experiments have demonstrated that \sys{} achieves great effectiveness for identifying faults in multi-language SAST tools.
There are some possible directions to further improve \sys{}.

First, as stated in Section~\ref{sec:methodology}, we currently use LLMs as the transformation engine to implement each metamorphic relation. 
In the future, we may consider using a multi-agent system instead of directly employing prompt-driven LLMs, as agents have long-term memory and tool-calling capabilities~\cite{agentsurvey}, which can help \sys{} generate more diverse variant programs, test deeper implementations of SAST tools, and accelerate the testing process. 
Moreover, a multi-agent system has strong task decomposition and scheduling abilities, making it a potential transformation selector for generating high-order variant programs that can identify faults more efficiently.

Second, similar to most testing research, \sys{} cannot completely avoid false positives, although our proposed heuristics can eliminate many of them. 
However, we can adopt a learning-based approach, for example, fine-tuning a large language model (LLM) to identify false positives and thereby improve the precision of \sys{}. 
Previous studies have demonstrated the potential of both traditional statistical learning and deep learning methods to reduce false positives~\cite{reducefp1, reducefp2}. 
By leveraging LLMs, with their large pre-trained corpora and strong capabilities in understanding program semantics and intentions of developers~\cite{reducefp3}, we can further enhance the effectiveness of false-positive reduction.

Third, future work could focus on enabling \sys{} to automatically discover and refine metamorphic relations, rather than relying solely on manually designed ones.
By mining large code repositories and analyzing historical bug patterns, \sys{} could learn novel transformation rules that capture domain-specific program behaviors.
Such an adaptive MR discovery mechanism would not only expand the diversity of generated program variants, but also improve the fault-revealing capability of metamorphic testing in previously unexplored scenarios.

\subsection{Threat to Validity}

\noindent \textbf{Internal Threat.}
The primary internal threat to validity stems from potential implementation bugs in \sys{}, the baseline approaches, and our experimental scripts.
Our key mitigation strategy is the public release of our testing framework, which facilitates verification and replication.
We are committed to addressing any issues identified through this open approach.

\noindent \textbf{External Threat.}
The main external threat lies in the evaluation targets themselves.
To demonstrate the generality of our approach, we apply \sys{} to six different SAAT tools across two programming languages.
Another threat stems from LLM hallucinations, which may generate incorrect variants that fail to satisfy the requirements specified in the prompts.
This issue is common to most LLM-based pipelines. Although we introduce a dynamic-feedback-driven mechanism to filter out some invalid variants, the remaining programs may still lead to false positives in our framework.
A third threat involves the differencing-based clustering approach.
Despite reducing the effort required to analyze experimental results, this remains a conservative strategy.
The unanalyzed variant programs may potentially harbor undetected faults in future evaluations.
Nevertheless, the number of faults already detected confirms the effectiveness of our approach, and we have further verified the uniqueness of faults within each cluster through random sampling.
\section{Related Work}

\subsection{Testing SAST Tools}
Several techniques have been proposed for testing SAST tools~\cite{2022icpcfinder, cuoq2012testing, klinger2019differentially, ase2024minotaur, ase2024sherlock, fse2024sabugs}.
Some previous work adopts differential testing to address the challenge of oracle design;
however, this approach depends on common components between two SAST tools, such as bug detectors~\cite{2022icpcfinder} and program representation construction algorithms~\cite{sascope}, which limits its scalability.
Some researches~\cite{stvr2017quickchecking, cgo2020testsa} are based on the specification, which requires manually and precisely formalizing the correct behavior of an analyzer, causing it to be hard and unrealistic.
Some use equivalent relations as the metamorphic relation to solve the lack of oracle problem in testing SAST tools~\cite{fse2021datalog, issta2023dependency, statfier, annatester}.
Similarly, we use equivalent relation (i.e., leveraging LLMs to generate variants in a semantics-preserving way and checking for differential analysis results of equivalent mutants) to address the oracle problem, but our work also can cover two novel types of metamorphic relations, increasing the semantics diversity of generated variants, and test SAST tools universally, not only limited to specific programming languages.

\subsection{LLM-based Testing Approaches}
The application of LLM-based approaches has been explored for testing diverse system software, including compilers~\cite{xia2024fuzz4all, wang2025mut4all}, deep learning frameworks~\cite{issta2025crossprobe, li2024enhancing}, and operating systems~\cite{tcad2024ecg}.
Currently, however, no research focuses on leveraging LLMs to test SAST tools.
Although LLM based compiler testing approaches~\cite{xia2024fuzz4all, mut4all} represent the most related domain, its methodologies cannot be directly transferred due to two primary reasons.
First, their test oracles are incompatible due to differing fault symptoms (particularly logic faults) between compilers and SAST tools.
For example, Fuzz4All~\cite{xia2024fuzz4all} focuses solely on crash bugs, and its oracle design is relatively simple.
Second, the program transformations designed for compiler testing are insufficient for comprehensively assessing SAST tools.
For instance, Fuzz4All only supports semantics-preserving transformations.
Our approach not only supports this type of transformation, but also enables program modifications through program repair and vulnerability injection.
Moreover, program transformations used in other domains usually cannot be directly applied to SAST tools due to grammar differences.

\subsection{SAST-related Study Work}
As a widely adopted technique in software security, several studies have investigated SAST tools~\cite{sast1, sast2, sast3, sast4}.
These studies have primarily focused on evaluating the performance and effectiveness of SAST tools~\cite{sast1, sast2}, examining how developers employ them in real-world scenarios~\cite{sast4}, or enhancing their detection capabilities through tool combination strategies~\cite{sast3}.
However, none of the existing work has explored the use of LLMs to assist software testing by systematically identifying and analyzing the faults of SAST tools.
Our work addresses this gap by leveraging LLMs to uncover potential weaknesses in SAST tools and provide targeted improvements, thereby enhancing their precision and reliability in practical security testing workflows.
\section{Conclusion}

We present \sys{}, a LLM-based universal testing framework that automatically generates input programs via semantic-preserving, program repair, and vulnerability injection transformations for discovering faults in SAST tools.
\sys{} leverage context information to synthesize prompts driving LLMs for variant generation and relies on two key heuristics, namely dynamic-feedback-driven variant selection and program-differencing-based clustering, to reduce false positives caused by the hallucination of LLMs and efforts to analyze evaluation results.
Our experiments show that \sys{} outperforms the existing researches by finding more bugs yet iterating through fewer variants.
Overall, \sys{} has discovered \faultcnt{} bugs, of which \fixedfaultcnt{} have been confirmed or fixed.
Our results confirms that developers of SAST tools supporting various programming languages can incorporate our approach into their test suite construction to further improve the tools' usability and reliability.
While we focus on faults that focused in the internal implementation of SAST tools in this paper, it would be worthwhile to study other types of faults in the future (e.g., inconsistencies between documentation and the behavior of SAST tools~\cite{tan2012tcomment}).
Another direction is to improve the effectiveness of \sys{} by tuning configurations of SAST tools as prior work shows that configurable software such as program verification tools and static analysis frameworks can be tuned~\cite{koc2021satune, icse2023ecstatic}.

\section*{Acknowledgments}
This work is supported by the Natural Sciences and Engineering Research Council of Canada (NSERC) Discovery Grants RGPIN-2024-04301.

% \section*{Data Availability} 
% The experimental data and source code are available at: https://llm4sast.github.io/.

\bibliographystyle{ACM-Reference-Format}
\bibliography{reference}

\end{document}